\documentclass[letterpaper,journal]{IEEEtran}
\usepackage[utf8]{inputenc}
\usepackage[english]{babel}
\IEEEoverridecommandlockouts
\usepackage[hyphens]{url}
\usepackage[hidelinks,hypertexnames=false]{hyperref}

\usepackage{cite}
\usepackage{mathptmx}
\usepackage{amsmath,amssymb,amsfonts}
\usepackage{graphicx}
\usepackage{textcomp}
\usepackage[table]{xcolor} 
\usepackage{multirow}
\usepackage{enumitem}
\usepackage{array}
\newcolumntype{x}[1]{>{\centering\arraybackslash\hspace{0pt}}p{#1}}
\usepackage{algorithm}
\usepackage[noend]{algpseudocode}
\definecolor{revcolor}{RGB}{0,84,166}
 
\definecolor{addcolor}{RGB}{0,0,200}
\definecolor{delcolor}{RGB}{180,0,0}
\definecolor{boxbg}{RGB}{240,248,255}

\usepackage{glossaries}
\glsdisablehyper
\newacronym{DOA}{DOA}{Direction-of-Arrival}
\newacronym{esprit}{ESPRIT}{estimation of signal parameters via rotational invariance techniques}
\newacronym{music}{MUSIC}{multiple signal classification}
\newacronym{rf}{RF}{radio frequency}
\newacronym{mr}{MR}{Moving Receiver}
\newacronym{imu}{IMU}{Inertial Measurement Unit}
\newacronym{tff}{TFF}{Transmitter at Far-Field}
\newacronym{uwb}{UWB}{Ultra-Wideband}
\newacronym{iot}{IoT}{Internet-of-Things}
\newacronym{DFT}{DFT}{Discrete Fourier Transform}
\newacronym{sa-esprit}{SA-ESPRIT}{single-antenna ESPRIT}
\newacronym{sa-music}{SA-MUSIC}{single-antenna MUSIC}
\newacronym{snr}{SNR}{signal-to-noise ratio}
\newacronym{vaa}{VAA}{Virtual Antenna Array}
\newacronym{rave}{RAVE}{Radio-Acoustical Virtual Environment}
\newacronym{tce}{TCE}{IEEE Transactions on Consumer Electronics}
\def\BibTeX{{\rm B\kern-.05em{\sc i\kern-.025em b}\kern-.08em
    T\kern-.1667em\lower.7ex\hbox{E}\kern-.125emX}}
\usepackage[left=1.62cm,right=1.62cm,top=1.9cm]{geometry}
 
\begin{document}
\frenchspacing
\hbadness=10000
\vbadness=10000
\hfuzz=3pt
 
\title{Lightweight Single-Antenna Direction-of-Arrival Estimation for
Curvilinear Trajectories in Mobile Embedded Systems}
 
\author{%
\IEEEauthorblockN{Nour Zidi\IEEEauthorrefmark{1},
  Jérémie Voix\IEEEauthorrefmark{2},
  Pascal Giard\IEEEauthorrefmark{1}, and
  Lokman Sboui\IEEEauthorrefmark{3}}
 
\IEEEauthorblockA{\IEEEauthorrefmark{1}Department of Electrical Engineering,
  \'Ecole de technologie sup\'erieure, Montr\'eal, Qu\'ebec, Canada}
 
\IEEEauthorblockA{\IEEEauthorrefmark{2}Department of Mechanical Engineering,
  \'Ecole de technologie sup\'erieure, Montr\'eal, Qu\'ebec, Canada}
 
\IEEEauthorblockA{\IEEEauthorrefmark{3}Department of System Engineering,
  \'Ecole de technologie sup\'erieure, Montr\'eal, Qu\'ebec, Canada}
 
\IEEEauthorblockA{Email: nour.zidi.1@ens.etsmtl.ca\quad
  \{jeremie.voix, pascal.giard, lokman.sboui\}@etsmtl.ca}
}
 
\maketitle
 
\begin{abstract}
Accurate direction-of-arrival (DOA) estimation is valuable for spatially selective communication in noisy industrial environments. This work investigates a lightweight single-antenna framework in which receiver motion forms a virtual aperture. The receiver uses onboard inertial measurement unit (IMU) headings and two-way-ranging (TWR) measurements to a known fixed beacon, avoiding GPS, optical tracking, and high-precision external tracking of the mobile receiver.

The curvilinear virtual-array MUSIC formulation is evaluated numerically using phase-coherent narrowband snapshots over arbitrary trajectories. Hardware experiments validate a range-domain TWR--IMU bearing estimator using corrected and averaged ranging observations. In a campaign of 100 consecutive four-revolution sweeps, all trials are retained. Phase-aligned accumulation recovers a 123~mm range-modulation amplitude, in close agreement with the measured 120~mm antenna lever arm. The single-sweep bearing precision is $8.2^\circ$; absolute world-frame accuracy is limited by systematic BNO055 magnetometer drift in the motorized setup. The embedded bearing-estimation computation consumes 144.9~mJ per estimate, approximately 3\% of the measured cycle energy in the tested configuration. These results establish the feasibility of onboard range-domain bearing estimation and motivate future experimental validation of phase-coherent MUSIC during free-form mobile trajectories.

\end{abstract}
 
\begin{IEEEkeywords}
Direction-of-Arrival (DOA), Curvilinear trajectory, Radio frequency (RF), Virtual Antenna Array (VAA), Multiple Signal Classification (MUSIC)
\end{IEEEkeywords}
\glsresetall
 
\section{Introduction}
\label{sec:intro}
 
\IEEEPARstart{T}{he} \gls{rave} framework is fundamentally an indoor technology, designed specifically for workers equipped with industrial hearing protection devices (HPDs) in high-noise environments such as
underground mines, manufacturing floors, and industrial warehouses~\cite{b15}. In these settings, the sky is typically not visible, GPS is unavailable, and the acoustic environment is so loud that workers cannot
communicate without electronic hearing protection. \gls{rave} addresses this by providing spatially-aware audio rendering inside the HPD: the wearer receives directionally adapted speech from co-workers without removing their hearing protection, preserving both safety and communication.

Even where GPS is available, metre-scale horizontal uncertainty is not sufficient to define a compact RF virtual array at a 6\,GHz carrier. The receiver-side framework avoids GPS and optical tracking of the mobile device while using TWR constraints from a known fixed beacon.

Traditional \gls{DOA} methods commonly rely on physical antenna arrays and must account for effects such as signal coherence and mutual coupling~\cite{b3}. Representative approaches include BLE-based \gls{DOA} estimation with phase-noise mitigation~\cite{b51}, nonlinear array interpolation for correlated signals~\cite{b8}, adaptive beamforming~\cite{b7}, fast planar-array estimation~\cite{b9}, single-moving-observer localization~\cite{b23}, phase-difference projection for non-uniform arrays~\cite{b10}, sparse off-grid and optimized coprime-array methods~\cite{b11,Sabbar2021}, and DFT-based virtual-array extension~\cite{b12}. While effective in specific scenarios, these methods remain constrained by their respective array geometries, phase ambiguities, grid errors, sensitivity to noise and model mismatch, mutual coupling, or computational cost.
 
Recent advances have demonstrated single-antenna receivers employing linear trajectories and low-cost hardware~\cite{b13}, and dual-antenna receivers in controlled motion~\cite{b14}. However, these techniques are restricted to linear or predetermined paths. Concurrently, deep learning and data-driven subspace methods~\cite{liu2018tap,elbir2020deepmusic,zhao2022cnn} have improved accuracy but remain constrained by significant computational demands. Coarray-tensor methods~\cite{b47}, hybrid dynamic-subarray architectures~\cite{b48}, and hybrid receiver architectures~\cite{b49} offer alternative high-resolution designs, though their predefined geometries and hardware configurations limit applicability to the unpredictable movements characteristic of RAVE scenarios. Coprime-array methods additionally require careful ambiguity handling and array-parameter selection~\cite{b50,Sabbar2021}. These limitations motivate lightweight, flexible \gls{DOA} frameworks for resource-constrained wearables and IoT devices.
 
These challenges are particularly critical in RAVE, where users navigate complex and unpredictable paths, and devices must function in real time under stringent hardware and energy constraints. Consistent with recent works on consumer indoor localization and direction-finding~\cite{b52,b53}, our research emphasizes a lightweight, embedded solution appropriate for battery-powered consumer electronics. The contribution of this work is primarily system-level rather than algorithmic:

\begin{enumerate}[leftmargin=*, label=(\arabic*)]
 
\item We formulate a single-moving-antenna virtual-array MUSIC model for arbitrary curvilinear trajectories and evaluate its sensitivity to SNR, snapshot count, source number, multipath, trajectory linearization, position uncertainty, aperture, and element count through numerical simulation.
\item We formulate a receiver-side trajectory framework using onboard IMU headings and TWR ranges to a known fixed anchor, without GPS, optical tracking, or motion-capture measurements of the receiver trajectory.
\item We validate the complete TWR--IMU measurement chain through controlled circular-trajectory experiments.
\item We measure the board-level energy of one complete ranging and bearing-estimation cycle on a 170\, MHz Cortex-M4 platform and discuss its relevance for GPS-denied RAVE deployments with a known fixed beacon.
\end{enumerate}

The rest of this paper is structured as follows. \autoref{sec:problem} introduces the problem formulation. \autoref{sec:proposed} outlines the proposed \gls{DOA} estimation framework. \autoref{sec:simulations} presents in-depth numerical simulations. \autoref{sec:experiments} validates the results with experiments. \autoref{sec:conclusion} encapsulates the main findings and future directions.

\emph{Notation}: Throughout the paper, $k=1,\ldots,K$ indexes sources, $m=1,\ldots,M$ indexes virtual array elements, $j=1,\ldots,M-1$ indexes trajectory segments, $i$ indexes raw TWR exchanges, and $n=1,\ldots,N_{\mathrm{snap}}$ indexes snapshots. Bold lowercase and uppercase symbols denote vectors and matrices, respectively; $(\cdot)^\top$ and $(\cdot)^H$ denote transpose and Hermitian transpose.

\section{Problem Formulation}
\label{sec:problem}
\subsection{System Model}
\label{sec:system_model}
 
\begin{figure}[t]
  \centering
  \includegraphics[width=\linewidth]{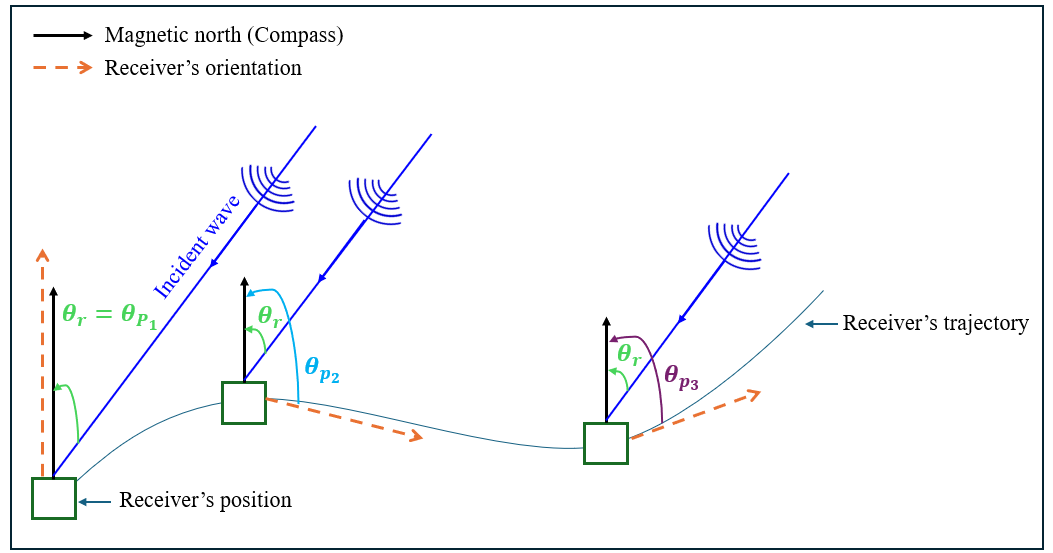}
  \caption{Illustration of \gls{DOA} estimation with a moving receiver
    equipped with a magnetic compass along a curvilinear trajectory.}
  \label{fig1}
\end{figure}

In this study, we consider the RAVE concept where \gls{DOA} is needed for audio adaptation with distance and angle. The system consists of a node with a single antenna that moves along a curvilinear trajectory, effectively creating a \gls{vaa}. For validation, we analyse the far-field case, where the beacon periodically sends communication signal packets. The transmitter position $\mathbf{r}_\text{tx}$ is assumed known in the base deployment, corresponding to a pre-surveyed infrastructure element or a self-reported beacon position. If the transmitter position is unknown, an additional bootstrapping phase using multi-hop TWR is required. To characterize the \gls{DOA}, as depicted in \autoref{fig1}, we define two angles: the first angle, $\theta_r$, is the azimuth between the direction of the incident wave (from the transmitter) and a fixed reference direction (magnetic north, as provided by the on-board magnetometer). The second angle, $\theta_{p,m}$, $m=1,2,\dots, M$, is the orientation of the receiver's antenna at the $m$-th position relative to the same reference. Denoting by $\theta_{\text{rel},m}$ the bearing of the incident wave measured in the antenna frame at position $m$, our primary goal is to estimate the reference angle $\theta_r$:
 
\begin{equation}
\theta_r = \theta_{\text{rel},m} + \theta_{p,m}.
\label{eq:theta_r}
\end{equation}
 
In the remainder of the paper, $\theta$ denotes a generic candidate bearing measured from this same magnetic-north reference; the simulation ground truth $\theta_0$ (\autoref{sec:simulations}) and the experimental anchor bearing (\autoref{sec:experiments}) are both instances of $\theta_r$, and $\hat\theta$ denotes its estimate.
 
The angle $\theta_{p,m}$ is tracked in real time by an \gls{imu} comprising accelerometers, gyroscopes, and a magnetometer. We use $L>\lambda/2$ as a practical aperture guideline ($2.5$\,cm at 6\,GHz): smaller apertures reduce directional phase diversity and degrade MUSIC conditioning and resolution, although they do not necessarily make the covariance rank-deficient. MUSIC also requires $M>K$ and sufficiently distinct steering vectors.
 
\subsection{Transmitter-Receiver Communication}
\label{sec:twr}
 
\begin{figure*}[t]
  \centering
  \includegraphics[width=0.9\textwidth]{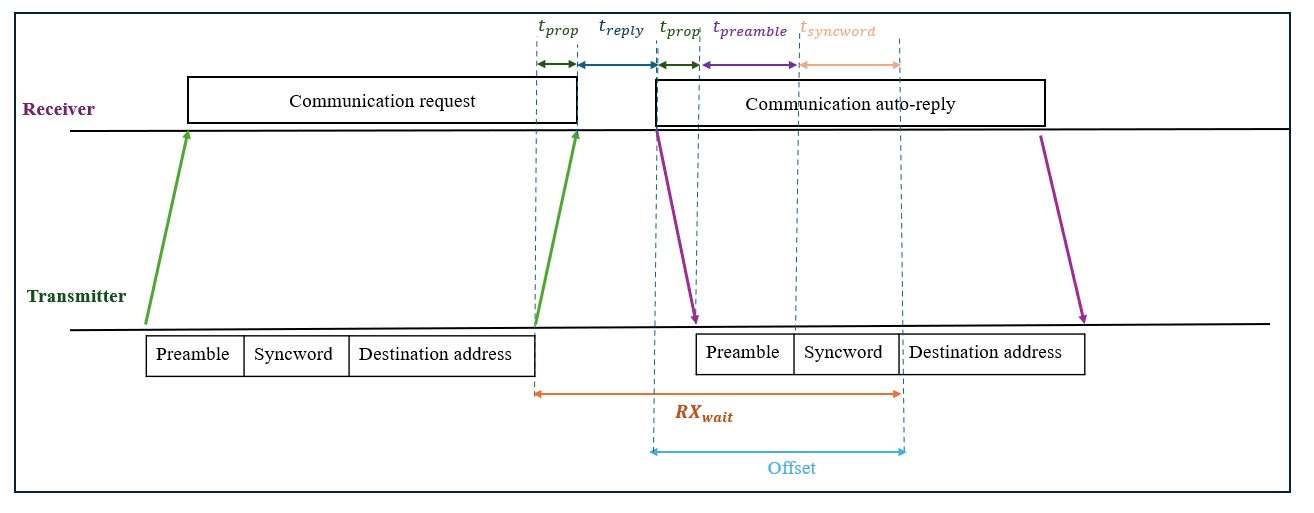}
  \caption{Communication process between transmitter and receiver.}
  \label{figComm}
\end{figure*}

To accurately determine the distance between a transmitter and a receiver, a two-way ranging (TWR) protocol is used as depicted in \autoref{figComm}. The objective is to measure the signal's round-trip propagation delay~\cite {b22}, which is translated into distance. The transmitter sends an initial request; the receiver replies with timing and phase information; the transmitter extracts both its own phase and the receiver's $RX_\text{wait}$ counter. A second exchange recovers the previously recorded phase. This exchange is repeated $N_\text{avg}$ times and averaged to increase measurement reliability.
 
\subsubsection{Offset Calculation}
\begin{equation}
  \text{Offset} = t_\text{reply} + t_\text{preamble} + t_\text{syncword}.
  \label{eq:offset}
\end{equation}
 
\subsubsection{Propagation Delay Calculation}
\begin{equation}
  t_\text{prop} = \frac{RX_\text{wait} - \text{Offset}}{2}.
  \label{eq:tprop}
\end{equation}
 
\subsubsection{Distance to Transmitter Calculation}
\begin{equation}
  \text{Distance} = c \times t_\text{prop}.
  \label{eqd}
\end{equation}

\subsubsection{TWR Noise Model}
\label{sec:twr_noise}

Let $\widehat{R}_i$ denote the round-trip counter value for the $i$th reported ranging result. It is modeled as
\begin{equation}
  \widehat{R}_i =
  \frac{2t_{\mathrm{prop},i}}{\Delta t_{\mathrm{clk}}}
  +\delta_{\mathrm{off}}
  +\epsilon_i^{(c)}+\epsilon_i^{(g)},
  \label{eq:twr_noise}
\end{equation}
where $t_{\mathrm{prop},i}$ is the one-way propagation delay, $\Delta t_{\mathrm{clk}}$ is the counter clock period, $\delta_{\mathrm{off}}$ is a static hardware offset, and $\epsilon_i^{(c)}$ and $\epsilon_i^{(g)}$ are the clock-quantization and residual timing-jitter errors, respectively, all expressed in counter cycles. The corresponding one-way distance estimate is
 
\begin{equation}
  \hat d_i =
  \frac{\widehat{R}_i-\delta_{\mathrm{off}}}{2}\,
  \Delta t_{\mathrm{clk}}\,c,
  \label{eq:d_hat}
\end{equation}
where $c$ is the speed of light.
 
Let $\sigma_{\mathrm{raw}}$ be the standard deviation of one single elementary TWR exchange. The ranging core averages $N_{\mathrm{avg}}$ exchanges to produce one reported distance value. Assuming independent elementary errors, the standard deviation of that reported distance is
 
\begin{equation}
  \sigma_{\mathrm{read}} \approx
  \frac{\sigma_{\mathrm{raw}}}{\sqrt{N_{\mathrm{avg}}}}.
  \label{eq:sigma_read}
\end{equation}
In the experiments, $\sigma_{\text{read}}$ is measured from stationary, post-averaging range samples, thereby including residual correlation and multipath. 
\subsubsection{Clock Synchronization}
\label{sec:clock_sync}
 
The SR1020 performs hardware clock synchronization during preamble acquisition, aligning each received packet to the initiator's symbol clock. The two-way ranging protocol inherently compensates for first-order clock offset errors, leaving residual drift negligible compared to the range uncertainty considered in this work.
 
\subsubsection{Range-Domain Observable and Scope of MUSIC Validation}
\label{sec:phase}
 
The SR1020 ranging interface provides timing-based observables from individual TWR exchanges. After static offset correction and averaging over $N_{\mathrm{avg}}$ exchanges, these measurements are converted into one-way range estimates using Eq.~\eqref{eq:d_hat}. Therefore, the retained data are range-domain observations rather than packet-to-packet phase-coherent complex baseband snapshots, which are not accessible from the used interface. Accordingly, the hardware estimator in \autoref{sec:experiments} exploits the measured range sequence and IMU headings to estimate the bearing of a known anchor. In the circular experiment, the bearing is recovered from the range-modulation phase described by Eq.~\eqref{eq:exp_range}. The MUSIC formulation in \autoref{sec:signal} is validated only through numerical simulations using phase-coherent snapshots and reconstructed curvilinear-array geometries. The experimental results validate the TWR--IMU measurement chain, range-domain angular estimation, embedded implementation, and energy consumption, but not carrier-phase-coherent MUSIC operation.

\section{Proposed DOA Estimation}
\label{sec:proposed}
 
\subsection{Virtual Element Position Estimation}
\label{sec:position}

Unlike conventional methods, the proposed approach does not rely on predefined or externally tracked positions; virtual-element coordinates are recovered entirely from on-board TWR and IMU. A single TWR measurement constrains the receiver to a circle of radius $\hat{d}_m$ centred at $\mathbf{r}_\text{tx}$ and does not yield a unique 2-D position. The combined estimator of Eq.~\eqref{eq:fusion} resolves this ambiguity by jointly minimizing:
 
\begin{equation}
  \min_{\{\mathbf{r}_m\}} \sum_{m=1}^{M}
    \bigl(\|\mathbf{r}_m-\mathbf{r}_\text{tx}\|_2 - \hat{d}_m\bigr)^2
  + \mu \sum_{m=2}^{M}
    \|\mathbf{r}_m - \mathbf{r}_{m-1} - \Delta\mathbf{r}_m\|_2^2,
  \label{eq:fusion}
\end{equation}
 
where the first term enforces the TWR range circle, the second enforces the IMU displacement $\Delta\mathbf{r}_m=\ell_{m-1}[\cos\alpha_{m-1},\sin\alpha_{m-1}]^\top$ along the $(m{-}1)$-th trajectory segment, 
 
Assuming independent zero-mean Gaussian measurement errors, a statistically
consistent choice of $\mu$ is obtained from the weighted least-squares
formulation as
 
\begin{equation}
\mu=
\frac{\sigma_{\mathrm{read}}^{2}}
{\sigma_{\Delta r}^{2}},
\label{eq:mu}
\end{equation}
 
where $\sigma_{\Delta r}$ is the standard deviation of the IMU-derived displacement increments.
 
The reconstruction is locally identifiable only under an initialized pose, nonzero motion increments, sufficiently divergent motion directions, and bounded range and heading errors. Non-collinearity of consecutive motion segments helps avoid locally degenerate geometries, but it does not by itself ensure global uniqueness.
The complete simulation workflow is summarized in Algorithm~\ref{alg1};its position-estimation stage involves:
\begin{enumerate}
  \item \emph{Distance Calculation}: at each element index $m$, the offset-corrected one-way TWR range \(d_m\) is obtained from the
reported round-trip counter.
  \item \emph{Position Estimation}: solve
    $\|\mathbf{r}_m-\mathbf{r}_\text{tx}\|=\hat d_m$ jointly with the IMU
    kinematic constraint via Eq.~\eqref{eq:fusion}.
  \item \emph{Trajectory Reconstruction}: repeat at each element index;
    estimated positions form the virtual array.
\end{enumerate}
 
\subsection{Signal Model and MUSIC Spectrum}
\label{sec:signal}
 
The time-domain signal model is:
\begin{equation}
  y_m(t) = \sum_{k=1}^K s_k(t)\,e^{-j2\pi f\tau_{m,k}(t)} + n_m(t),
  \label{eqy}
\end{equation}
where $s_k(t)$ is the $k$-th source signal, $n_m(t)$ is additive noise, and
the delay $\tau_{m,k}(t)=d_{m,k}/c$ with
$d_{m,k}=\sqrt{(r_{x,m}-x_k)^2+(r_{y,m}-y_k)^2}$.

In the far field, with $\mathbf{u}(\theta_k)=[\cos\theta_k,\sin\theta_k]^\top$ the unit vector pointing from the array origin towards the $k$-th source and $\rho_k$ the source--origin distance,
\begin{equation}
  d_{m,k}\approx \rho_k - \mathbf{u}^\top(\theta_k)\,\mathbf{r}_m,
  \qquad \mathbf{r}_m=[r_{x,m},r_{y,m}]^\top.
\end{equation}
 
The common bulk term $\rho_k$ contributes an identical phase to all elements and is dropped, so the steering vector of the curvilinear virtual array at a candidate direction $\theta$ is the $M\times1$ column vector
\begin{equation}
  \mathbf{a}(\boldsymbol{\alpha},\theta) =
  \exp\!\bigl(\,j k_w\,
  \mathbf{R}_M^\top(\boldsymbol{\alpha})\,\mathbf{u}(\theta)\bigr),
  \label{eq:sv}
\end{equation}
where $k_w=2\pi/\lambda$ and the exponential is applied element-wise to the $M\times 1$ vector of projected positions
$\mathbf{R}_M^\top(\boldsymbol{\alpha})\,\mathbf{u}(\theta)$. MUSIC requires $M>K$ and a nondegenerate direction-sensitive aperture: the steering vectors and their directional derivatives must vary across the virtual-element positions. This condition is assessed explicitly in the numerical simulations.

The MUSIC pseudo-spectrum is:
\begin{equation}
  P_\text{MUSIC}(\theta)=
  \frac{1}{\mathbf{a}^H(\boldsymbol{\alpha},\theta)\,
  \mathbf{E}_n\mathbf{E}_n^H\,
  \mathbf{a}(\boldsymbol{\alpha},\theta)},
  \label{eq:music}
\end{equation}
where $\mathbf{E}_n$ spans the noise subspace of $\widehat{\mathbf{R}}_{yy}=\frac{1}{N_{\mathrm{snap}}}
\sum_{n=1}^{N_{\mathrm{snap}}}\mathbf{y}(t_n)\mathbf{y}^H(t_n)$ (the $M-K$ eigenvectors corresponding to its smallest eigenvalues), and DOAs are identified from peaks of $P_\text{MUSIC}$~\cite{b16,b17,b18}. Forming the sample covariance requires $\mathcal{O}(M^2N_{\mathrm{snap}})$ operations, while eigendecomposition requires $\mathcal{O}(M^3)$ operations; the small virtual arrays considered here keep both costs practical.
 
\begin{algorithm}[t]
\caption{Offline simulation workflow for phase-coherent curvilinear VAA--MUSIC}
\label{alg1}
\begin{algorithmic}[1]
\State \textbf{Input:} true trajectory $\mathbf R_{\mathrm{true}}$;
known anchor position; simulated TWR observations, IMU headings and motion
increments; coherent narrowband source signals; $K$
\State Generate noisy TWR and IMU observations from $\mathbf R_{\mathrm{true}}$
\State Reconstruct virtual-element coordinates
$\widehat{\mathbf R}_{\mathrm{TWR+IMU}}$ using Eq.~\eqref{eq:fusion}
\State Generate phase-coherent complex snapshots using
$\mathbf R_{\mathrm{true}}$ and Eq.~\eqref{eqy}
\State Form $\widehat{\mathbf R}_{yy}=N_{\mathrm{snap}}^{-1}
\sum_{n=1}^{N_{\mathrm{snap}}}\mathbf y[n]\mathbf y^H[n]$ 
\State Construct each MUSIC steering vector using
$\widehat{\mathbf R}_{\mathrm{TWR+IMU}}$
\State Compute $P_{\mathrm{MUSIC}}(\theta)$ via Eq.~\eqref{eq:music}
\State Identify DOA peaks and compare them with the true source bearings
\State \textbf{Output:} MUSIC DOA errors under TWR--IMU-induced
array-manifold mismatch
\end{algorithmic}
\end{algorithm}
 
\subsection{Curvilinear Motion Linearization}
\label{sec:linearization}
 
\begin{figure}[t]
  \centering
  \includegraphics[width=0.9\linewidth]{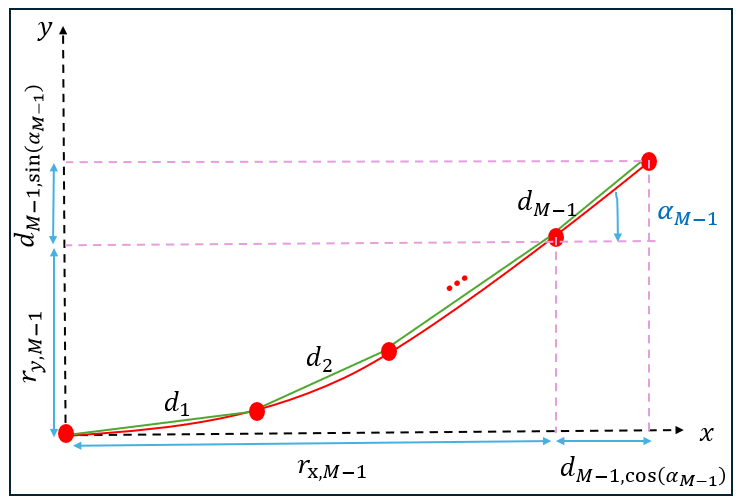}
  \caption{Trajectory in the 2-D curved plane.}
  \label{fig2}
\end{figure}

The curvilinear trajectory is approximated by piecewise-linear segments, as shown in \autoref{fig2}. 
Let $\mathbf{r}_m^\text{true}=[r_{x,m}^\text{true},\,r_{y,m}^\text{true}]^\top$ denote the true position of the $m$-th virtual antenna, where $m=1,\ldots,M$, $\ell_m$ is the inter-sample distance between the $m$-th and $(m{+}1)$-th samples, and $\alpha_m$ is the local heading angle of the $m$-th segment, $m=1,\ldots,M-1$. Then
\begin{equation}
  r_{x,m+1} = r_{x,m} + \ell_m\cos\alpha_m,\quad
  r_{y,m+1} = r_{y,m} + \ell_m\sin\alpha_m.
  \label{eq7}
\end{equation}

\paragraph{Linearization error bound.}
Let $\hat{\mathbf{r}}_m$ be the linearized approximation of $\mathbf{r}_m^\text{true}$, let $\kappa_{\max}$ denote the maximum
curvature along the trajectory, and let $\ell_{\max}=\max_j \ell_j$ be the largest segment length. The linearization error satisfies
\begin{equation}
  \bigl\|\hat{\mathbf{r}}_m - \mathbf{r}_m^\text{true}\bigr\|
  \leq \frac{\kappa_{\max}\,\ell_{\max}^2}{8},
  \label{eq:lin_bound}
\end{equation}
 
where $\|\cdot\|$ is the Euclidean norm.

The coordinates of the $m$-th virtual antenna are approximated as
\begin{equation}
  r_{x,m}\approx\sum_{j=1}^{m-1}\ell_j\cos\alpha_j,\quad
  r_{y,m}\approx\sum_{j=1}^{m-1}\ell_j\sin\alpha_j,
  \label{eq8}
\end{equation}
where $\ell_j$ is the inter-sample distance of the $j$-th segment and $\alpha_j$ is its corresponding heading angle. This gives the full coordinate matrix
 
\begin{equation}
  \mathbf{R}_M(\boldsymbol{\alpha})=
  \begin{bmatrix}
    0, & r_{x,2}, & \cdots, & r_{x,M}\\
    0, & r_{y,2}, & \cdots, & r_{y,M}
  \end{bmatrix},
\end{equation}
where $\boldsymbol{\alpha}=[\alpha_1,\alpha_2,\ldots,\alpha_{M-1}]^\top$
collects all segment heading angles.
 
\subsection{Range-Domain Geometric Bearing-Error Model}
\label{sec:pos_error_model}
 
A first-order uncertainty model is derived for the range-domain bearing estimator used in the circular experiment. The model is specific to the range-based estimator and is not a bound for phase-coherent MUSIC, whose geometry sensitivity is evaluated separately in \autoref{sec:sens_sim}.
 
Let $\bar{\mathbf p}_m\in\mathbb{R}^{2}$ denote the centered nominal position of virtual element $m$,
\begin{equation}
\sum_{m=1}^{M}\bar{\mathbf p}_m=0 ,
\end{equation}
where $M$ is the number of virtual elements. The position uncertainty is modeled as
\begin{equation}
\delta\mathbf p_m=
\mathbf b+\beta\mathbf J\bar{\mathbf p}_m+
\boldsymbol{\varepsilon}_m ,
\qquad
\mathbf J=
\begin{bmatrix}
0&-1\\
1&0
\end{bmatrix},
\end{equation}
where $\mathbf b$ represents a common translation, $\beta$ a small heading bias, and
\begin{equation}
\boldsymbol{\varepsilon}_m\sim
\mathcal N(0,\sigma_{\mathrm{pos}}^2\mathbf I_2)
\end{equation}
is the differential coordinate error.
 
For a source located at range $\rho$ and bearing $\theta$, with
\[
\mathbf u(\theta)=
[\cos\theta,\sin\theta]^\top ,
\qquad
\mathbf u'(\theta)=
[-\sin\theta,\cos\theta]^\top ,
\]
the range observation is approximated by
\begin{equation}
d_m\simeq
\rho-\mathbf u^\top(\theta)\bar{\mathbf p}_m+e_m ,
\end{equation}
 
where $e_m$ denotes range-readout error with standard deviation $\sigma_{\mathrm{read}}$. Combining independent readout and differential position errors gives
\begin{equation}
\sigma_{\mathrm{eff}}^2=
\sigma_{\mathrm{read}}^2+\sigma_{\mathrm{pos}}^2.
\end{equation}
 
The centered virtual-array geometry is characterized by
\begin{equation}
\mathbf S=
\frac1M\sum_{m=1}^{M}
\bar{\mathbf p}_m\bar{\mathbf p}_m^\top ,
\end{equation}
and its cross-range aperture is
\begin{equation}
W(\theta)=
\left(
\mathbf u'^{\top}(\theta)\mathbf S\mathbf u'(\theta)
\right)^{1/2}.
\label{eq:W}
\end{equation}
The resulting first-order bearing uncertainty is therefore
\begin{equation}
\sigma_{\hat{\theta}}=
\frac{\sigma_{\mathrm{eff}}}
{\sqrt{M}\,W(\theta)}.
\label{eq:B1}
\end{equation}
The bound is defined for $W(\theta)>0$, corresponding to a trajectory with nonzero cross-range aperture relative to the source direction.
\begin{figure}
  \centering
  \includegraphics[width=\linewidth]{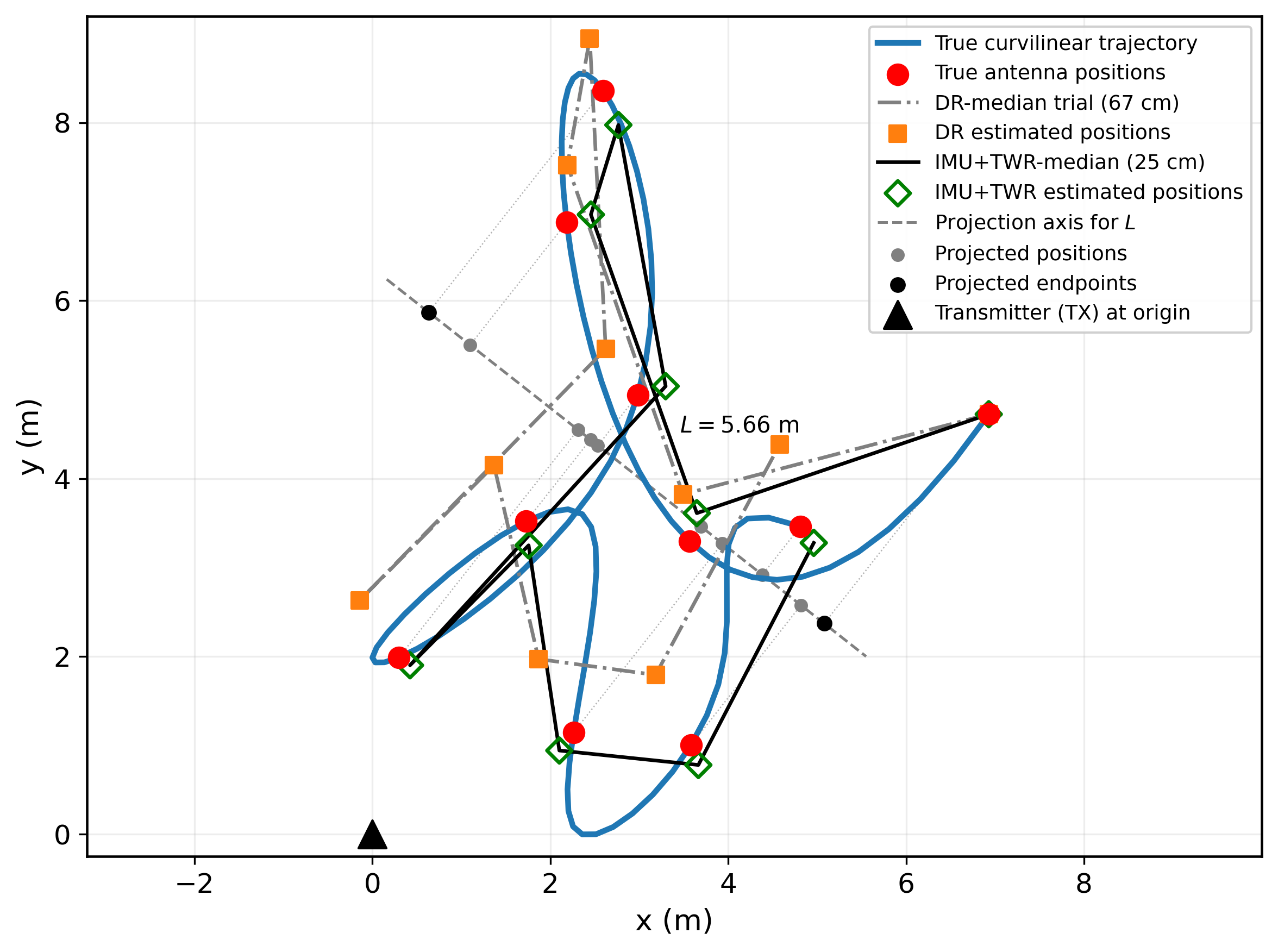}
 
\caption{Position reconstruction for a large-scale random curvilinear trajectory ($N_\mathrm{MC}=1000$ trials, $\sigma_\mathrm{read}\approx 151\,\mathrm{mm}$, $N_\mathrm{avg}=128$ exchanges per session); the median trial is shown for dead reckoning (DR) and for the fused IMU+TWR estimator. Gray markers indicate the antenna positions projected onto the cross-range axis (perpendicular to the transmitter bearing, dashed gray line), whose extent defines the effective aperture $L = 5.66$\,m relevant to the cross-range dispersion of Eq.~\eqref{eq:W}.}
  \label{fig3}
\end{figure}

\section{Numerical Simulations}
\label{sec:simulations}
 
\begin{figure*}[t]
  \centering
  \includegraphics[width=0.9\linewidth]{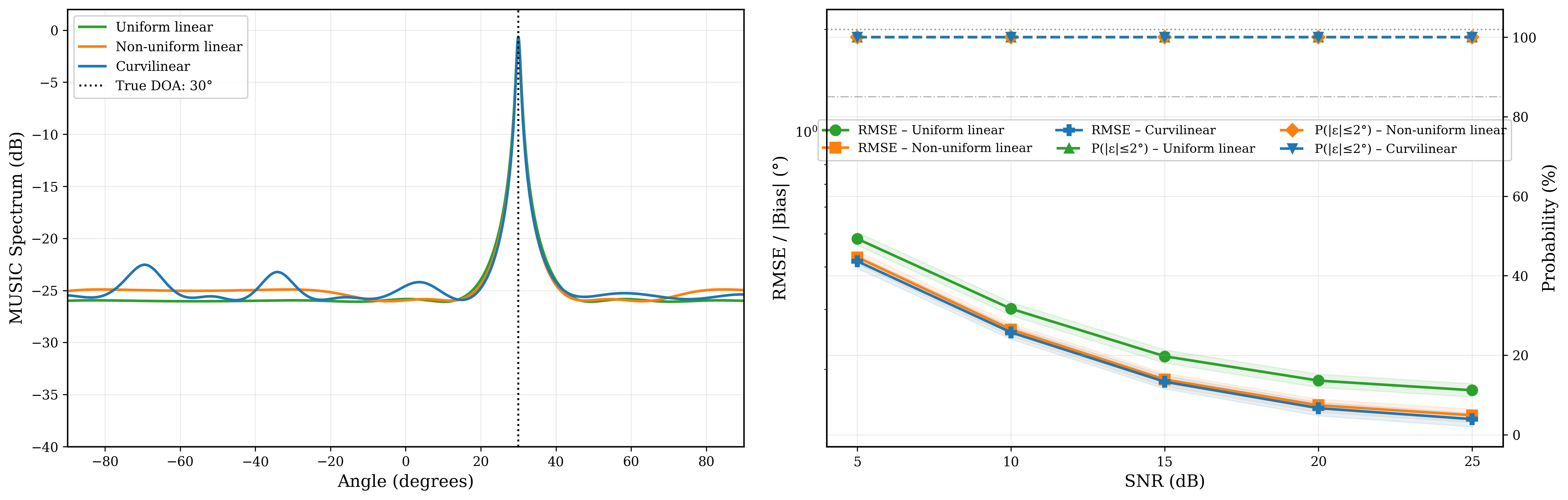}
  \caption{Influence of SNR on MUSIC DOA estimation for three trajectory types at $f=6$\,GHz, using a compact virtual antenna array with approximately $3.5\lambda$ maximum aperture, $M=6$, $N_{\mathrm{snap}}=16$, and $N_{\mathrm{MC}}=1000$. \emph{Left}: Monte-Carlo-averaged MUSIC pseudo-spectra at SNR$=15$\,dB. \emph{Right}: DOA RMSE (solid curves, left axis) and
  $P(|\varepsilon|\leq2^\circ)$ (dashed curves, right axis) versus SNR.
  Shaded regions denote bootstrap 95\% confidence intervals for the RMSE curves.}
  \label{fig4}
\end{figure*}
 
The proposed DOA framework is evaluated through numerical simulations using a narrowband far-field source at $\theta_0=30^\circ$ and $f=6$\,GHz ($\lambda=5$\,cm). Unless otherwise stated, simulations use $M=6$ virtual elements, $N_{\mathrm{snap}}=16$ snapshots, and $N_{\mathrm{MC}}=1000$ Monte-Carlo trials with complex AWGN.
 
The compact curvilinear VAA is generated from ordered points along a smooth trajectory, centred and scaled to a maximum aperture of approximately $3.5\lambda$ ($17.5$\,cm). Confidence intervals correspond to bootstrap 95\% bounds or $\pm1\sigma$ variations as indicated in the figure captions.
 
For MUSIC evaluation, phase-coherent snapshots are generated using the true trajectory $\mathbf R_{\mathrm{true}}$, while the steering vectors are constructed from the reconstructed geometry $\widehat{\mathbf R}_{\mathrm{TWR+IMU}}$. The resulting array-manifold mismatch quantifies the impact of trajectory reconstruction errors on coherent virtual-array processing.
 
Trajectory reconstruction is driven by the measured SR1020 ranging noise, so that the reconstruction study reflects the observed hardware behaviour. Injecting the measured single-shot ranging noise $\sigma_{\mathrm{raw}}\approx1.7$\,m (Section~\ref{sec:ranging_error}) and averaging over $N_{\mathrm{avg}}=128$ exchanges gives the range-readout uncertainty
\[
\sigma_{\mathrm{read}}
=\frac{\sigma_{\mathrm{raw}}}{\sqrt{N_{\mathrm{avg}}}}
\approx151~\mathrm{mm},
\]
consistent with the value measured directly from static captures in Section~\ref{sec:ranging_error}. Under this model, the fused IMU+TWR trajectory estimator reduces the median element-position error from 67\,cm (dead reckoning) to 25\,cm over an approximately 8\,m trajectory (\autoref{fig3}). The effect of residual geometry uncertainty is further studied in \autoref{sec:sens_sim}.
 
A $2^\circ$ DOA error threshold is used as an application-level reference, and both RMSE and $P(|\varepsilon|\leq2^\circ)$ are reported when applicable.
 
\subsection{Influence of Signal-to-Noise Ratio}
\label{sec:snr}
 
\autoref{fig4} evaluates the influence of SNR from 5 to 25\,dB. The left panel shows Monte-Carlo-averaged MUSIC pseudo-spectra at SNR$=15$\,dB. For all three trajectory types, the dominant peak is centred near the true source direction of $30^\circ$.
 
The right panel shows that the RMSE decreases as SNR increases. The success probability $P(|\varepsilon|\leq2^\circ)$ remains above 85\% over the evaluated SNR range and approaches 100\% at high SNR for all three trajectory types. At SNR$=5$\,dB, the MUSIC peak broadens because of noise; above approximately 10\,dB, the three trajectories show similar peak sharpness and estimation performance. These results support the use of curvilinear virtual-array motion for DOA estimation at 6\, GHz.
 
\subsection{Influence of the Number of Snapshots}
\label{sec:snapshots}
 
\begin{figure}[t]
  \centering
  \includegraphics[width=0.9\linewidth]{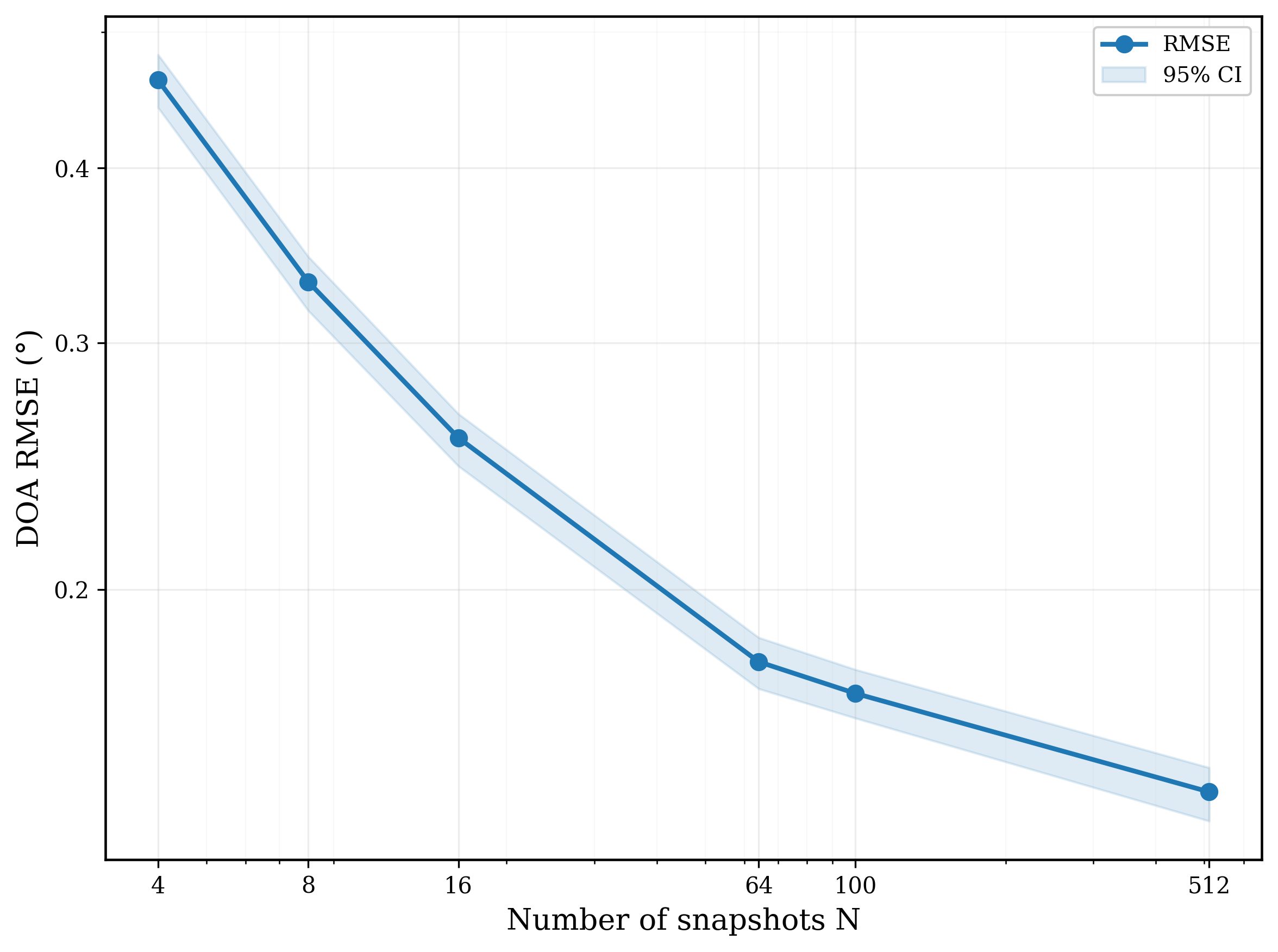}
  \caption{DOA RMSE versus snapshot count
  $N_{\mathrm{snap}}\in\{4,8,16,64,100,512\}$ at SNR$=10$\,dB,
  $f=6$\,GHz, $M=6$, $\theta_0=30^\circ$, and
  $N_{\mathrm{MC}}=1000$. Shaded regions denote bootstrap 95\%
  confidence intervals.}
  \label{fig7}
\end{figure}
 
As shown in \autoref{fig7}, increasing $N_{\mathrm{snap}}$ improves sample-covariance estimation, reducing DOA RMSE and narrowing the bootstrap confidence intervals. 
 
\subsection{Influence of the Number of Sources}
\label{sec:sources}
 
\begin{figure}[t]
  \centering
  \includegraphics[width=0.9\linewidth]{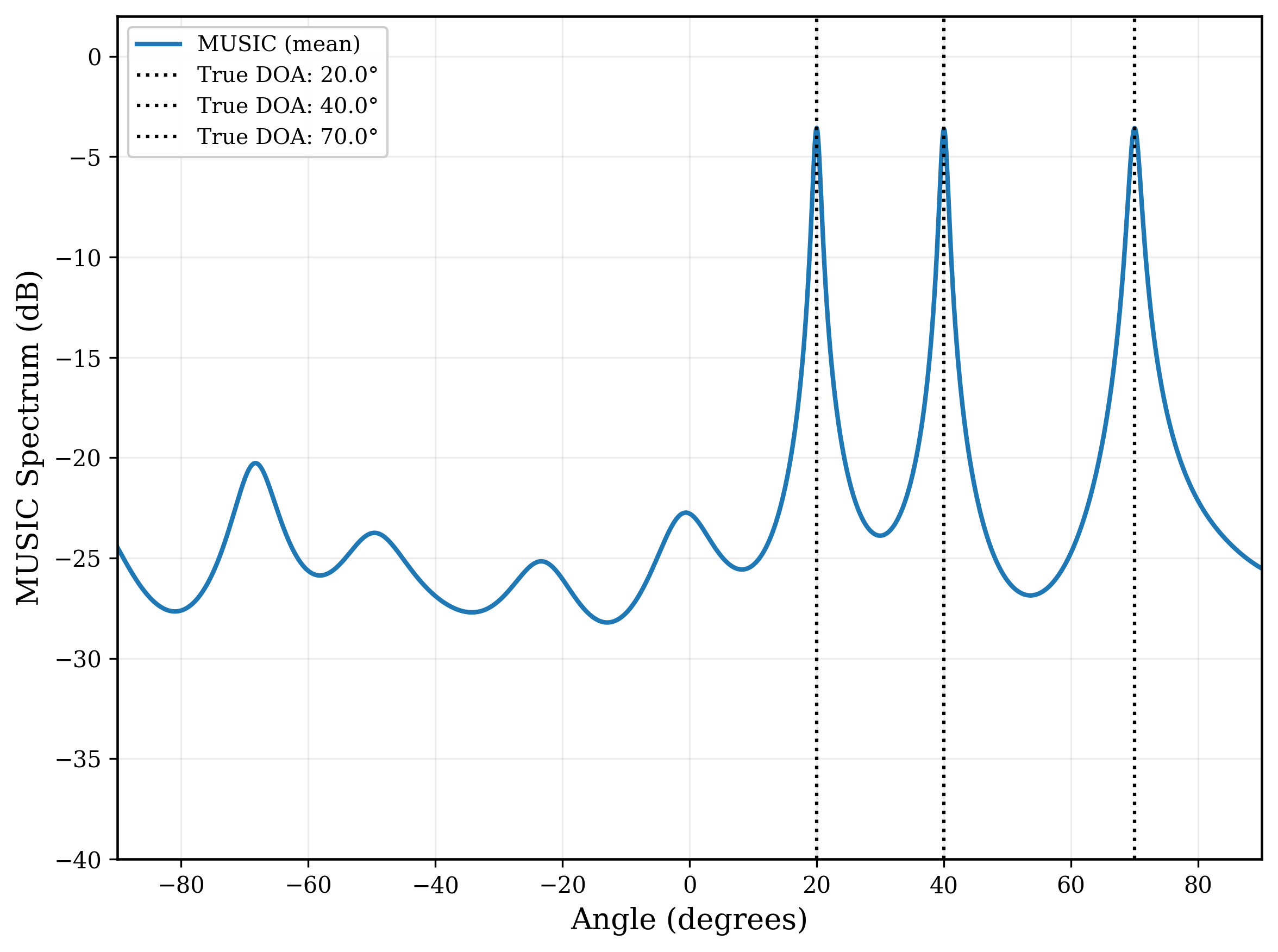}
  \caption{MUSIC pseudo-spectrum for three sources at $20^\circ$, $40^\circ$, and $70^\circ$, using the curvilinear VAA at $f=6$\,GHz, SNR$=10$\,dB, $N_{\mathrm{snap}}=64$, and $N_{\mathrm{MC}}=1000$.}
  \label{fig8_sources}
\end{figure}
 
\autoref{fig8_sources} illustrates the MUSIC pseudo-spectrum for three simultaneous sources at $20^\circ$, $40^\circ$, and $70^\circ$. The MUSIC noise subspace is constructed using the known source count $K=3$. The resulting pseudo-spectrum exhibits peaks near the three simulated source directions, demonstrating multi-source estimation for the tested angular separations.

\subsection{Multipath Robustness and Outlier Gating}
\label{sec:multipath_sim}
 
\begin{figure}[t]
  \centering
  \includegraphics[width=0.9\linewidth]{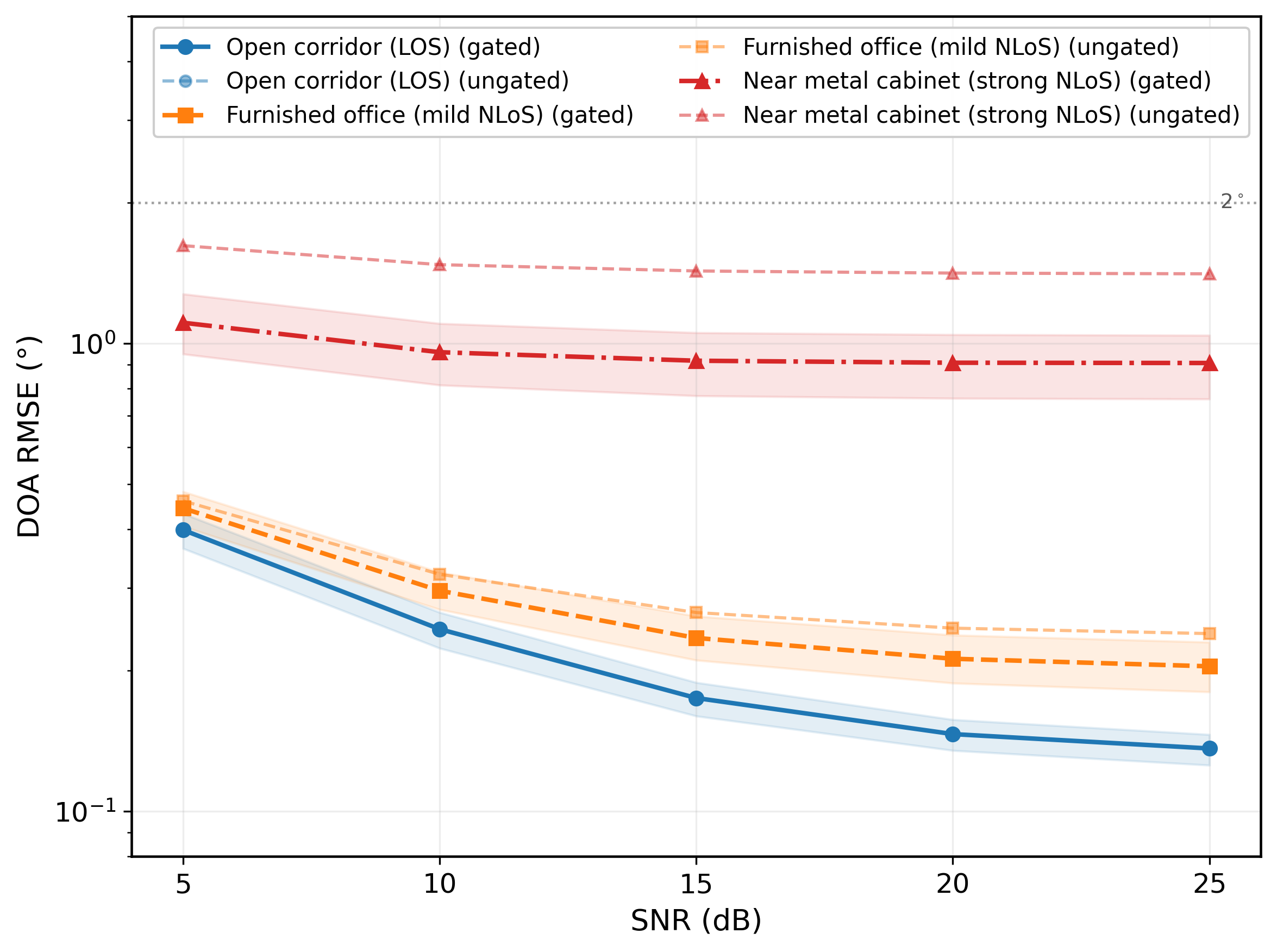}
  \caption{DOA RMSE versus SNR for three simulated propagation conditions at $f=6$\,GHz, $M=6$, $N_{\mathrm{snap}}=16$, $\theta_0=30^\circ$, and $N_{\mathrm{MC}}=1000$. Solid curves use SR1020 range-domain outlier gating with
  \texttt{CLOCK\_CYCLE\_COMPARE}$=0.5$; dashed curves omit the gate. The dotted horizontal line denotes the $2^\circ$ RMSE reference.}
  \label{fig9}
\end{figure}
 
\autoref{fig9} evaluates the effect of range-domain outlier gating under three simulated propagation conditions. Range observations identified as inconsistent by the SR1020 filtering rule are discarded before averaging and trajectory reconstruction. The resulting reduction in virtual-element geometry error improves the subsequent MUSIC estimate.
 
The simulated conditions are:
\begin{enumerate}[leftmargin=*]
  \item \emph{Open corridor (LOS):} no reflected component is included.
  \item \emph{Furnished office (mild NLoS):} a reflected component is
  present in 30\% of trials, with relative gain 0.4 and $20^\circ$ angular
  offset.
  \item \emph{Near metal cabinet (strong NLoS):} a reflected component is
  present in 70\% of trials, with relative gain 0.7 and $15^\circ$ angular
  offset.
\end{enumerate}

For the LOS condition, gating has a negligible effect because no corrupted range observations are simulated. In the mild-NLoS condition, gating reduces the RMSE from $0.35^\circ$ to $0.27^\circ$ at SNR$=15$\,dB. In the strong-NLoS condition, it reduces the RMSE from approximately $1.7^\circ$ to $0.9^\circ$; these results are summarized in \autoref{tab:multipath_sim}. Across the simulated trials, $\xi_1/\xi_2>6$\,dB, where $\xi_1$ and $\xi_2$ are the largest and second-largest eigenvalues of the sample covariance matrix $\widehat{\mathbf{R}}_{yy}$; therefore, no spatial-rank collapse is observed in this six-element VAA configuration.
 
\begin{table}[t]
\centering
\caption{DOA RMSE at SNR$=15$\,dB for three simulated propagation
conditions ($N_{\mathrm{MC}}=1000$, $M=6$, and $\theta_0=30^\circ$).}
\label{tab:multipath_sim}
\small
\setlength{\tabcolsep}{3pt}
\begin{tabular}{|p{2.9cm}|c|c|c|}
\hline
\textbf{Environment} &
\textbf{With gating} &
\textbf{Without gating} &
\textbf{Reduction} \\
\hline
Open corridor (LOS)              & 0.20 & 0.20 & 0\,\%  \\
Furnished office (mild NLoS)     & 0.27 & 0.35 & 23\,\% \\
Near metal cabinet (strong NLoS) & 0.90 & 1.70 & 47\,\% \\
\hline
\end{tabular}
\end{table}
 
\subsection{Linearization Error Verification}
\label{sec:lin_sim}

\autoref{tab:lin_error} summarizes the linearization-induced DOA error for different curvature $\kappa$ and inter-sample distance $\ell$ combinations.For each geometry, the maximum element-position displacement predicted by Eq.~\eqref{eq:lin_bound} is applied to the virtual-array coordinates. Signals are generated using the nominal geometry, whereas MUSIC evaluates the displaced geometry. This isolates trajectory-linearization error as an array-manifold mismatch.
 
Within the simulated operating envelope($\kappa\leq2$\,rad/m and $\ell\leq10$\,cm), the mean DOA error does not exceed $0.36^\circ$. The corresponding position displacement is at most 2.5\, mm, well below the heuristic reference $\sigma^*=\lambda/4=1.25$\, cm at 6\, GHz. Errors exceed $2^\circ$ only at $\ell=20$\, cm with $\kappa\geq1$\,rad/m, which corresponds to coarse sampling under strongly curved motion.
 
\begin{table}[t]
\centering
\caption{Mean linearization-induced DOA error ($^\circ$) versus curvature $\kappa$ and inter-sample distance $\ell$
($N_{\mathrm{MC}}=1000$, SNR$=15$\,dB, $f=6$\,GHz, and $M=6$). The operating envelope is $\kappa\leq2$\,rad/m and $\ell\leq10$\,cm. Bold entries exceed $2^\circ$.}
\label{tab:lin_error}
\begin{tabular}{|c|cccc|}
\hline
\multirow{2}{*}{$\kappa$ (rad/m)} &
\multicolumn{4}{c|}{$\ell$ (cm)} \\
& 2 & 5 & 10 & 20 \\
\hline
0.5 & 0.16 & 0.07 & 0.05 & 0.27 \\
1.0 & 0.16 & 0.08 & 0.08 & \textbf{2.10} \\
2.0 & 0.16 & 0.10 & 0.36 & \textbf{15.99} \\
3.0 & 0.17 & 0.13 & 0.47 & \textbf{32.12} \\
5.0 & 0.18 & 0.20 & 1.40 & \textbf{49.94} \\
\hline
\end{tabular}
\end{table}
 
\subsection{Speed Sensitivity}
\label{sec:speed}
 
Over the evaluated 0.5--3~m/s range, traversal speed had only a modest effect on DOA RMSE when the virtual-array geometry was reconstructed at each measurement position.
 
\subsection{Impact of Position Error on DOA Estimation}
\label{sec:sens_sim}
 
\begin{figure}[t]
  \centering
  \includegraphics[width=0.9\linewidth]
  {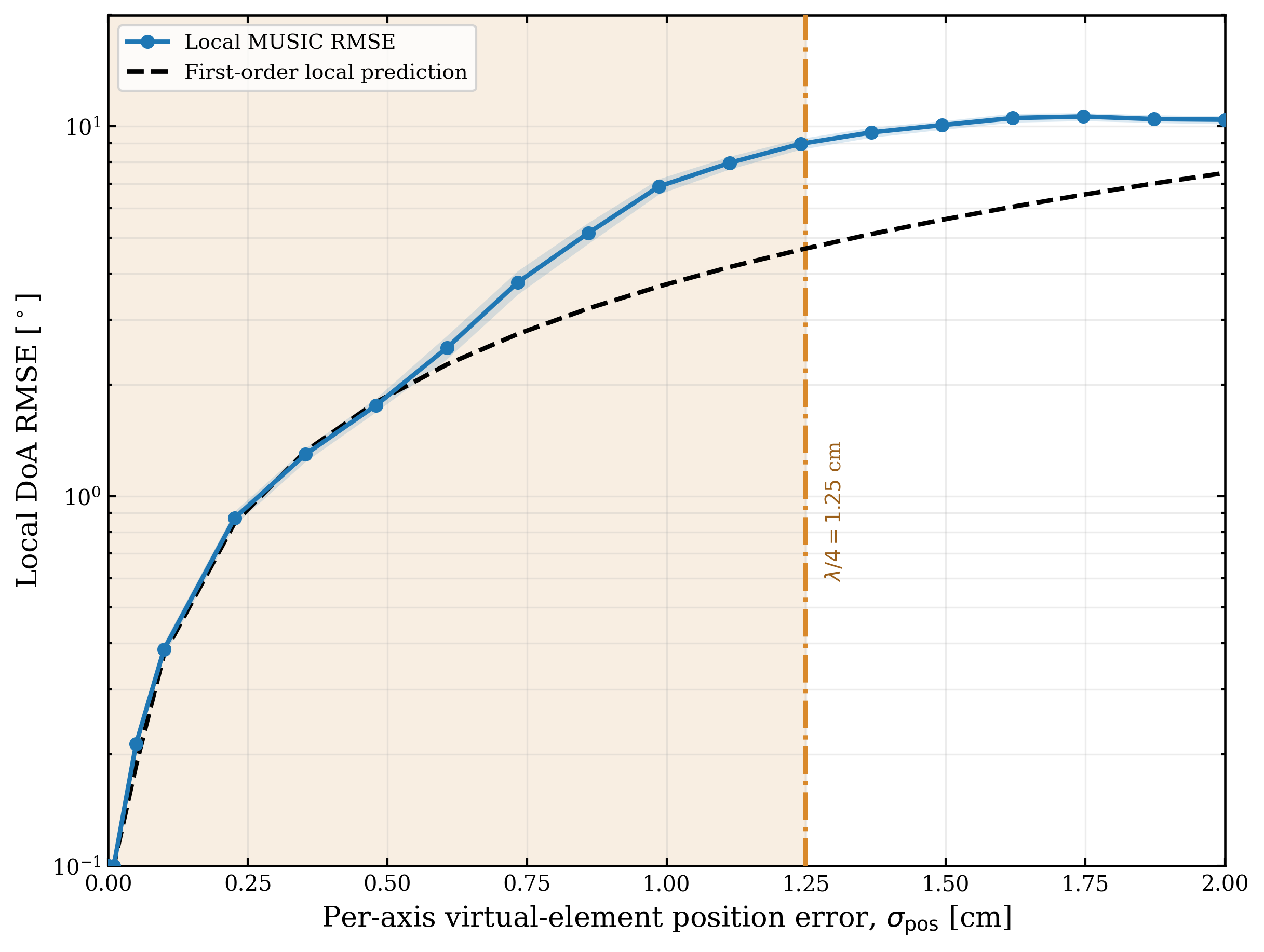}
  \caption{Local MUSIC DOA RMSE versus per-axis virtual-element position uncertainty for the compact curvilinear VAA at 6\,GHz
  (SNR$=15$\,dB, $N_{\mathrm{snap}}=16$, and $N_{\mathrm{MC}}=1000$). Blue markers denote the Monte-Carlo RMSE. The dashed black curve shows the first-order local prediction. The orange dash-dot line indicates the
  heuristic reference $\sigma^*=\lambda/4=1.25$\,cm.}
  \label{fig_sens}
\end{figure}
 
To assess trajectory-reconstruction uncertainty, independent zero-mean Gaussian perturbations with per-axis standard deviation $\sigma_{\mathrm{pos}}$ model the residual differential error of the TWR--IMU reconstructed coordinates. The received data are generated using the true geometry, whereas MUSIC constructs its steering vectors using the perturbed reconstructed geometry. This isolates reconstruction-induced array-manifold mismatch.
 
\autoref{fig_sens} shows that, at small position errors, the Monte-Carlo RMSE follows the first-order local prediction. At larger $\sigma_{\mathrm{pos}}$, the empirical RMSE exceeds the linearized prediction, reflecting increasing array-manifold mismatch. The $\lambda/4$ line is a wavelength-scaled heuristic reference, not a strict failure threshold. Accurate trajectory reconstruction is therefore required to maintain reliable DOA estimation.
 
\subsection{Resolution and Element-Count Scaling}
\label{sec:resolution}
 
\begin{figure}[t]
  \centering
  \includegraphics[width=0.9\linewidth]{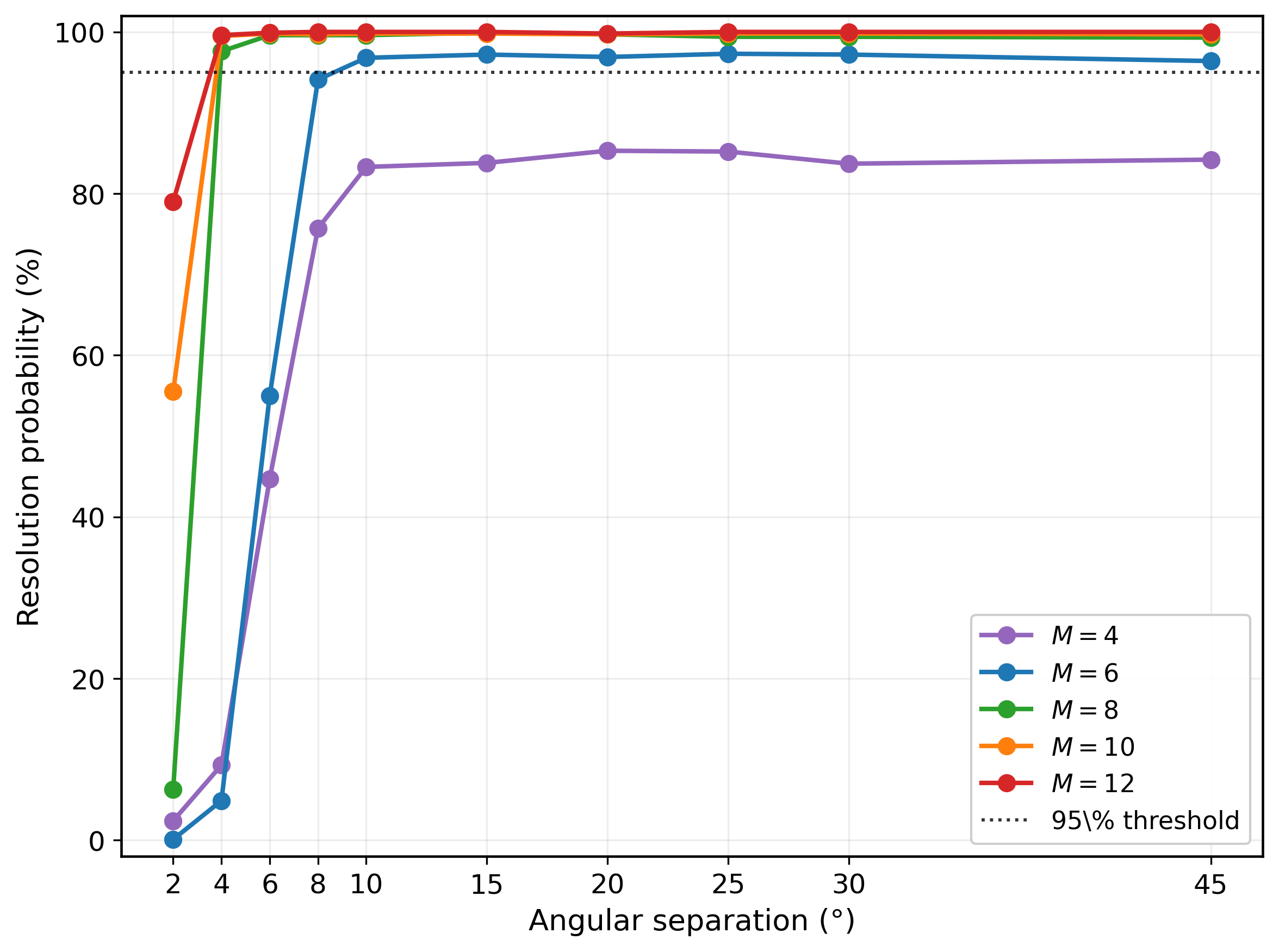}
  \caption{Two-source resolution probability versus angular separation for
  $M\in\{4,6,8,10,12\}$ virtual elements at $f=6$\,GHz,
  SNR$=15$\,dB, $N_{\mathrm{snap}}=16$, and
  $N_{\mathrm{MC}}=1000$. The horizontal dashed line
  denotes the 95\% resolution criterion.}
  \label{fig8_res}
\end{figure}
 
\autoref{fig8_res} evaluates two-source resolution probability as a function of angular separation. A trial is considered resolved when MUSIC produces two distinct pseudo-spectrum peaks associated with the two true source directions.
 
For $M\geq6$, the resolution probability exceeds 95\% for angular separations of approximately $8^\circ$ and above. Arrays with $M\geq8$ reach the same criterion at approximately $4^\circ$, demonstrating the benefit of increasing the number of virtual elements. In contrast, $M=4$ saturates near 85\% even at larger separations, indicating insufficient spatial information for the adopted 95\% reliability criterion.
 
\subsection{Comparison with Baseline Methods}
\label{sec:baseline}
 
\begin{figure}[t]
  \centering
  \includegraphics[width=0.8\linewidth]{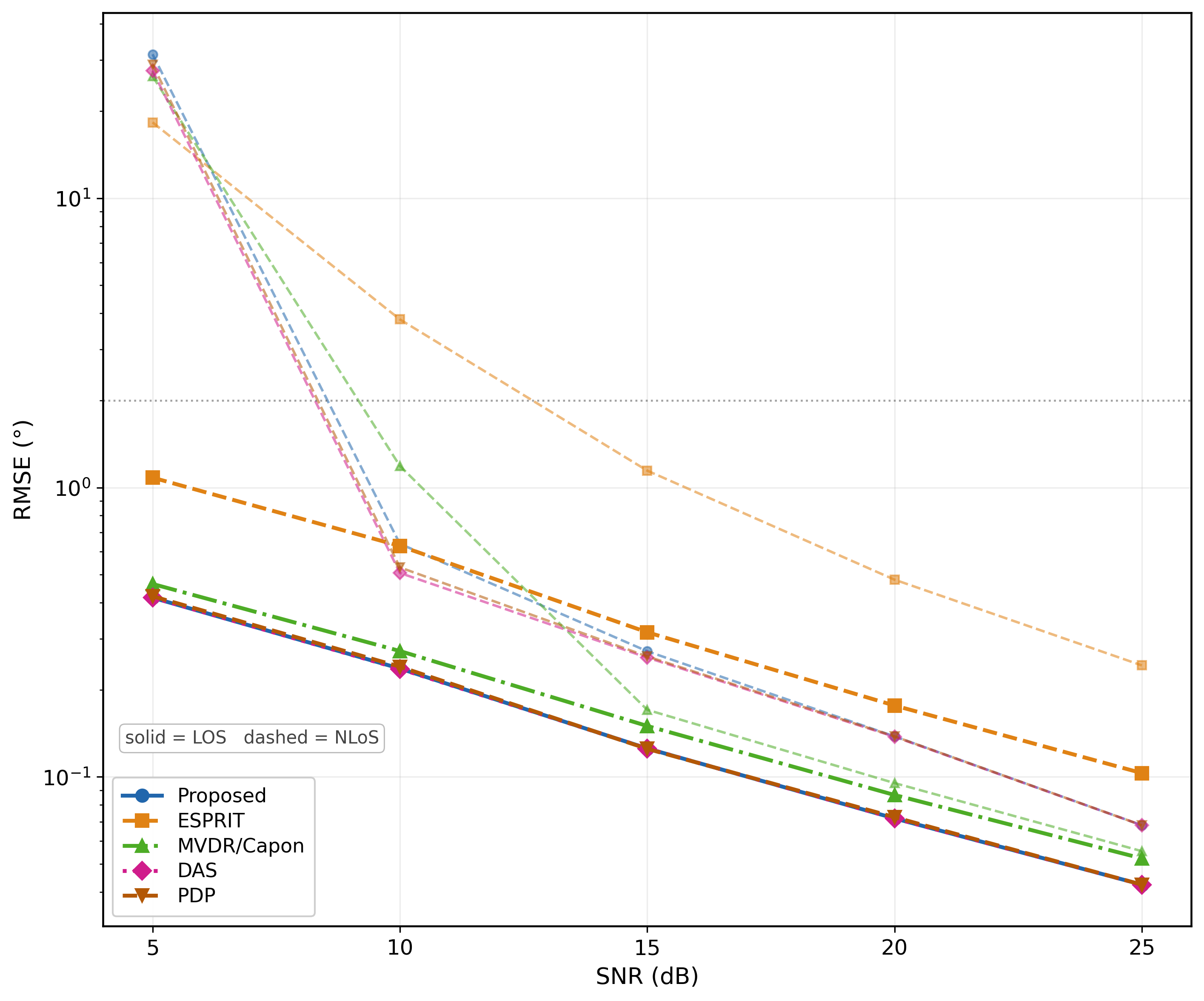}
  \caption{DOA RMSE versus SNR for the proposed curvilinear-VAA MUSIC method and four baseline estimators. Solid and dashed curves denote LOS and incoherent-interference conditions, respectively; interference power is 6\,dB above the noise floor. Parameters are $N_{\mathrm{MC}}=1000$, $M=6$, $N_{\mathrm{snap}}=16$, $\theta_0=30^\circ$, and $f=6$\,GHz. The dotted line denotes the $2^\circ$ RMSE reference.}
  \label{fig_baseline_bar}
\end{figure}

A Monte-Carlo comparison with $N_{\mathrm{MC}}=1000$ trials per configuration is performed against ESPRIT, MVDR/Capon, DAS, and PDP. All methods are evaluated at $f=6$\,GHz with $M=6$, $N_{\mathrm{snap}}=16$, and a true DOA of $30^\circ$. \autoref{fig_baseline_bar} reports DOA RMSE versus SNR for LOS and incoherent interference conditions, while \autoref{tab:comparison} summarizes the principal implementation characteristics of the methods.
 
In LOS, the proposed curvilinear-VAA MUSIC method achieves RMSE comparable to DAS and PDP over the evaluated SNR range. MVDR/Capon approaches similar accuracy at high SNR, whereas ESPRIT has a larger RMSE for the simulated geometry. Under incoherent interference, the proposed method maintains low RMSE at moderate and high SNR, while ESPRIT and MVDR/Capon are more strongly affected. The primary benefit of the proposed approach is its compatibility with a single moving antenna and a reconstructed curvilinear VAA, rather than LOS accuracy alone.
 
\begin{table}[t]
\centering
\scriptsize
\setlength{\tabcolsep}{1pt}
\caption{Qualitative characteristics of the evaluated DOA estimators. Quantitative RMSE results are reported in \autoref{fig_baseline_bar}.}
\label{tab:comparison}
\begin{tabular}{|p{1.35cm}|c|c|c|c|}
\hline
Method & Array model & Complexity & Array requirement &
VAA support \\
\hline
\textbf{Proposed} & Curvilinear & Low & 1 moving antenna & Yes \\
\hline
ESPRIT \cite{b6} & Shift-invariant & Medium & $\geq2$ elements & No \\
\hline
MVDR /Capon \cite{capon1969} & General known geometry & Medium &
$\geq2$ elements & Yes$^\dagger$ \\
\hline
DAS \cite{vantrees2002} & General known geometry & Low &
$\geq2$ elements & Yes$^\dagger$ \\
\hline
PDP \cite{b10} & Non-uniform linear & Low--Medium &
$\geq2$ elements & No \\
\hline
\end{tabular}
 
\vspace{1mm}
\footnotesize{$^\dagger$Requires reconstructed virtual-element coordinates; the method does not itself reconstruct the trajectory.}
\end{table}
\section{Experimental Results with UWB Radios}
\label{sec:experiments}
This section validates the embedded TWR--IMU range-domain bearing estimator on hardware. The phase-coherent MUSIC estimator of \autoref{sec:signal} is validated in simulation; the hardware campaign evaluates the range-domain estimator (the phase of a rotation-induced range modulation) together with the ranging noise and heading behaviour that govern its accuracy. Three findings structure the results: the SR1020 ranging noise is characterized and averages predictably; the compact rotating aperture produces a range modulation that, although near the noise level on a single sweep, is recovered by phase-aligned accumulation; and each sweep yields a bearing estimate with an inferred precision of $8.2^\circ$, while the absolute world-frame reference is limited by magnetometer drift in the motorized setup.
 
\subsection{Hardware Setup and Measurement Model}
\begin{figure}[t]
    \centering
    \includegraphics[width=0.6\linewidth]{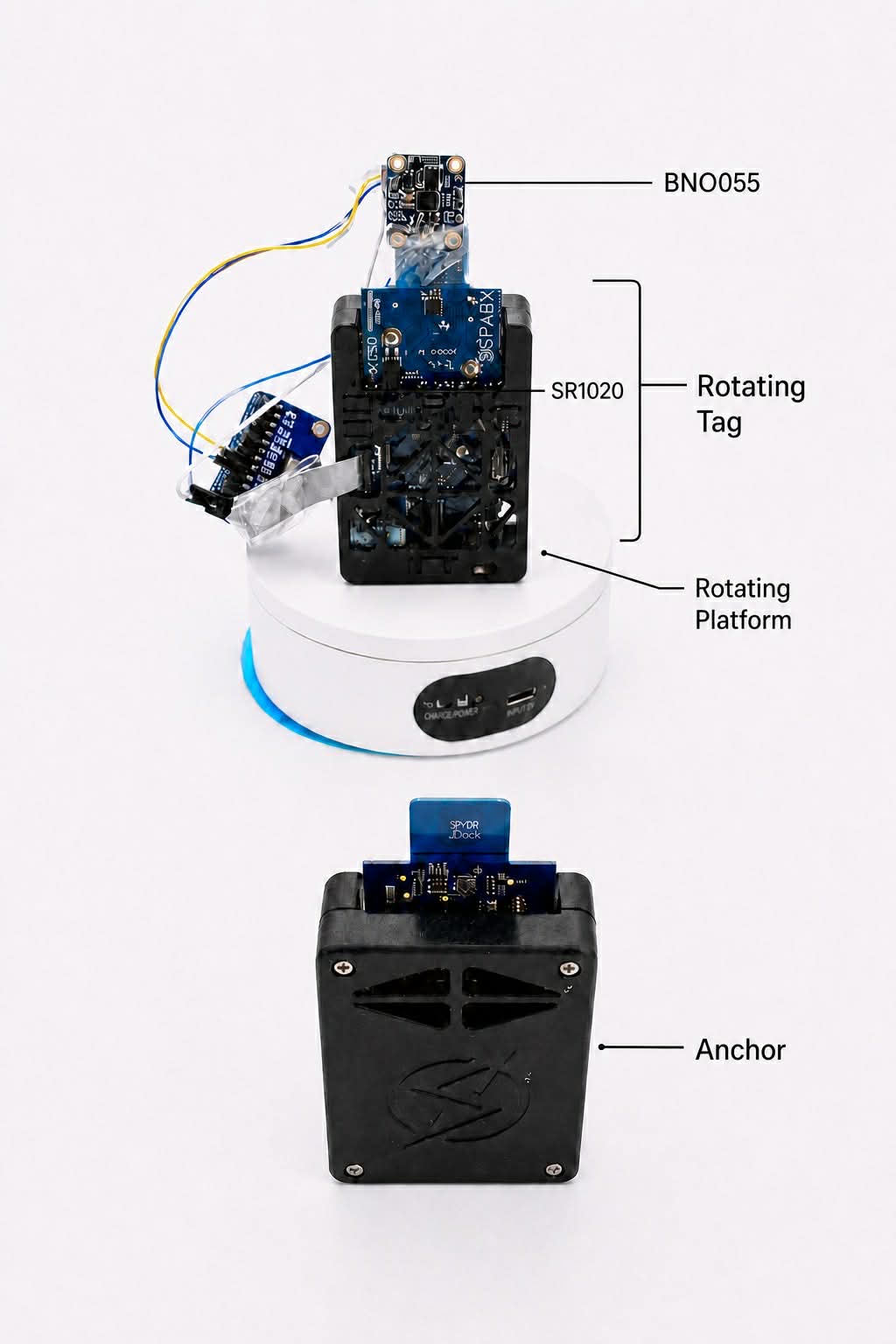}
    \caption{Hardware setup: the rotating tag (top), comprising an SR1020 UWB transceiver and a BNO055 IMU mounted on a motorized rotating platform at a known 120~mm antenna lever arm, and the fixed anchor (bottom), a second SR1020 transceiver used as the known reference node for two-way ranging.}
    \label{fig:setup}
\end{figure}
The setup illustrated in \autoref{fig:setup} uses two SPARK Microsystems SR1020 UWB transceiver boards operating in the 6.0--9.3\,GHz band~\cite{b25}: a fixed anchor and a tag mounted on a motorized rotating platform. The tag integrates a BNO055 \gls{imu}, and the complete bearing-estimation firmware runs on a 170\,MHz Arm Cortex-M4 (STM32G473). The tag antenna is mounted at a radial lever arm $r = 120$\,mm from the rotation axis forming a compact circular virtual aperture of diameter $2r$ (approximately$4.8\lambda$ at 6\,GHz). Each trial consists of one continuous motorized sweep of four revolutions.

The tag and anchor antennas were maintained in the same nominal vertical orientation, and the tag rotated about the vertical axis, preserving nominal linear-polarization alignment. No polarization or calibrated antenna-pattern correction was applied; arbitrary wearable orientation, body blockage, antenna tilt, and polarization mismatch are therefore outside the present scope.
 
The SR1020 ranging core reports one distance per $N_{\mathrm{avg}} = 128$ raw TWR exchanges; at the 500\,Hz exchange rate this corresponds to an averaging window $T_{\mathrm{avg}} = 256$\,ms. Because the reported distance corresponds to the centre of this window while the cached \gls{imu} heading is read later, each heading is corrected as
\begin{equation}
\phi_{\mathrm{corr}} = \phi_{\mathrm{cache}}
+ \omega\!\left(t_{\mathrm{age}} - \tfrac{T_{\mathrm{avg}}}{2}\right),
\label{eq:heading_time_correction}
\end{equation}
where $\omega$ is the estimated signed rotation rate. Time-aligned observations are grouped into $M = 12$ angular bins. A bin is counted as populated when it contains at least 4 valid observations, and a sweep is retained when at least 8 of the 12 bins are populated. Actual bin occupancy varies because of invalid-range rejection and nonuniform angular sampling. All 100 sweeps met the retention criteria. The range observed at heading $\phi_m$ follows
\begin{equation}
d(\phi_m)\approx d_0 - r\cos(\phi_m-\theta) + e_m \quad (r \ll d_0),
\label{eq:exp_range}
\end{equation}
so that rotation maps the anchor bearing $\theta$ to the phase of a sinusoidal range modulation of amplitude $r$. A nonlinear least-squares fit of Eq.~\eqref{eq:exp_range} recovers the bearing from the modulation phase.
 
\subsection{Ranging-Error Characterization}
\label{sec:ranging_error}
 
\begin{figure}[t]
\centering
\includegraphics[width=\linewidth]{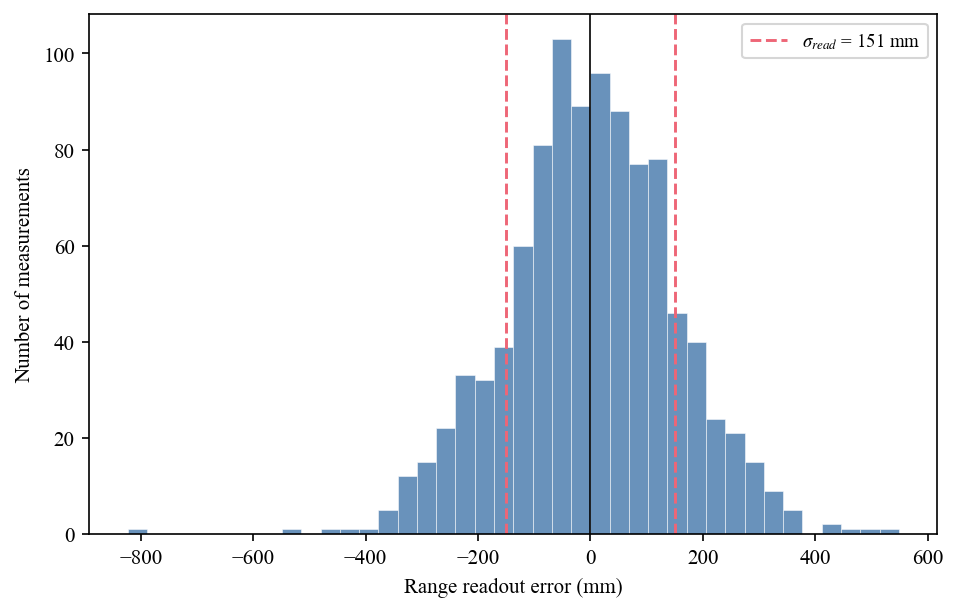}
\caption{Distribution of 1000 averaged range readings at a fixed 5\,m distance
($N_{\mathrm{avg}} = 128$), giving a range-readout standard deviation
$\sigma_{\mathrm{read}} \approx 151$\,mm (readout RMSE $\approx 151$\,mm).}
\label{fig:ranging_readout}
\end{figure}

The ranging error was measured from static calibration captures of 1000 averaged range readings at fixed anchor--tag distances, so that any observed dispersion reflects measurement noise rather than geometric variation. After $N_{\mathrm{avg}} = 128$ averaging, the range-readout standard deviation is $\sigma_{\mathrm{read}} \approx 151$\,mm at 5\,m (\autoref{fig:ranging_readout}), with a corresponding readout RMSE of $151$\,mm; at 1.3\,m it is $\approx 198$\,mm. The uncertainty remains of the same order at both distances, suggesting that distance alone is not the dominant dispersion source. Reducing the averaging to $N_{\mathrm{avg}} = 32$ raises $\sigma_{\mathrm{read}}$ to $\approx 303$\,mm, a factor of $2.01$ relative to the $N = 128$ case, matching the $\sqrt{128/32} = 2$ prediction and confirming that the readout noise averages as $1/\sqrt{N}$.

These values characterize the controlled, nominally polarization-aligned setup. During rotation, the 29.8\,s capture interval provides approximately 116 reported range observations before validity filtering, corresponding to about 9--10 observations per angular bin on average for \(M=12\). The actual occupancy varies because of invalid-range rejection and nonuniform angular sampling. The measured per-bin RMSE is $\approx136$\,mm because it also includes correlated ranging error, heading misalignment, multipath, and finite-bin variation; thus the static $1/\sqrt{N}$ law does not directly predict this residual.
 
\subsection{Range Modulation and Effective Aperture}
\label{sec:range_modulation}
 
\begin{figure}[t]
\centering
\includegraphics[width=\linewidth]{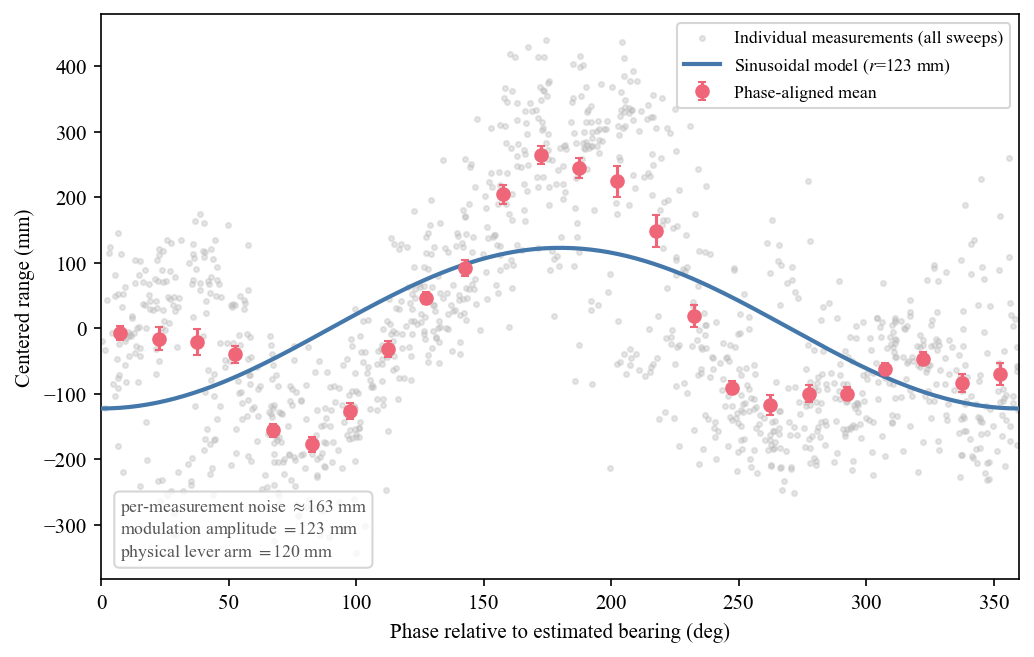}
\caption{Phase-aligned average of the centered range observations across the 100-sweep campaign. The recovered modulation amplitude ($123$\,mm) matches the measured physical lever arm ($120$\,mm), confirming the range-domain model despite a single-sweep amplitude SNR near unity.}
\label{fig:phase_aligned}
\end{figure}

On a single sweep, the $120$\,mm range modulation is of the same order as the per-bin ranging noise, so the sinusoid is only partially visible (\autoref{fig:phase_aligned}). Averaging the range observations across all sweeps, aligned by their estimated bearing, recovers the underlying modulation: the phase-aligned mean follows the expected $-r\cos(\phi - \theta)$ law with a fitted amplitude of $123$\,mm, in close agreement with the independently measured physical lever arm of $120$\,mm. This cross-validation confirms both the range-domain measurement model and the geometric origin of the modulation.
 
\subsection{Bearing-Estimation Accuracy}
\label{sec:doa_results}
 
The estimator is evaluated over 100 consecutive automated sweeps at a fixed anchor; all sweeps are retained. The world-frame anchor bearing is referenced through the static IMU heading with the tag antenna aimed at the anchor, read at the start of the campaign ($\theta_{\mathrm{ref}} = 104.0^\circ$).
 
Each sweep produces a complete DOA estimate from the sinusoidal fit over the full rotation. Two effects govern its accuracy. First, the absolute anchor bearing, which should be constant across sweeps, exhibits a systematic drift of $2.0^\circ$ per sweep (approximately $200^\circ$ over the campaign) originating in the BNO055 magnetometer, perturbed by the rotating metallic platform and motor; this drift affects the absolute heading reference, not the range-domain estimator. Second, to quantify the estimator precision independently of this slow drift, we use the standard deviation of the DOA change between consecutive sweeps. A nearly constant drift contributes primarily to the mean of the successive differences rather than to their standard deviation. After removing the mean successive increment, the remaining sweep-to-sweep variation yields a single-sweep DOA precision of $\sigma_\Delta/\sqrt{2} = 8.2^\circ$, consistent with the near-unity per-bin amplitude SNR and without requiring a parametric drift model. The magnetometer drift, by contrast, is a correlated systematic error that a single heading sensor cannot remove; suppressing it in deployment is precisely the role of the inertial fusion and multi-anchor geometry of the RAVE framework.
 
\begin{table}[t]
\centering
\scriptsize
\caption{Experimental bearing-estimation summary over the 100-sweep campaign
(fixed anchor).}
\label{tab:doa_stats}
\begin{tabular}{|l|c|}
\hline
\textbf{Metric} & \textbf{Value} \\
\hline
Number of sweeps & 100 \\
Physical lever arm $r$ & $120$\,mm \\
Recovered modulation amplitude & $123$\,mm \\
Readout noise $\sigma_{\mathrm{read}}$ (5\,m) & $151$\,mm \\
Per-bin ranging RMSE & $136$\,mm \\
Per-bin amplitude SNR & $\approx 0.9$ \\
Single-sweep DOA precision & $8.2^\circ$ \\
Magnetometer drift & $2.0^\circ$/sweep \\
\hline
\end{tabular}
\end{table}
\subsection{Power, Energy, and Battery Life}
\label{sec:power}
\begin{figure*}[t]
\centering
\includegraphics[width=0.85\linewidth]{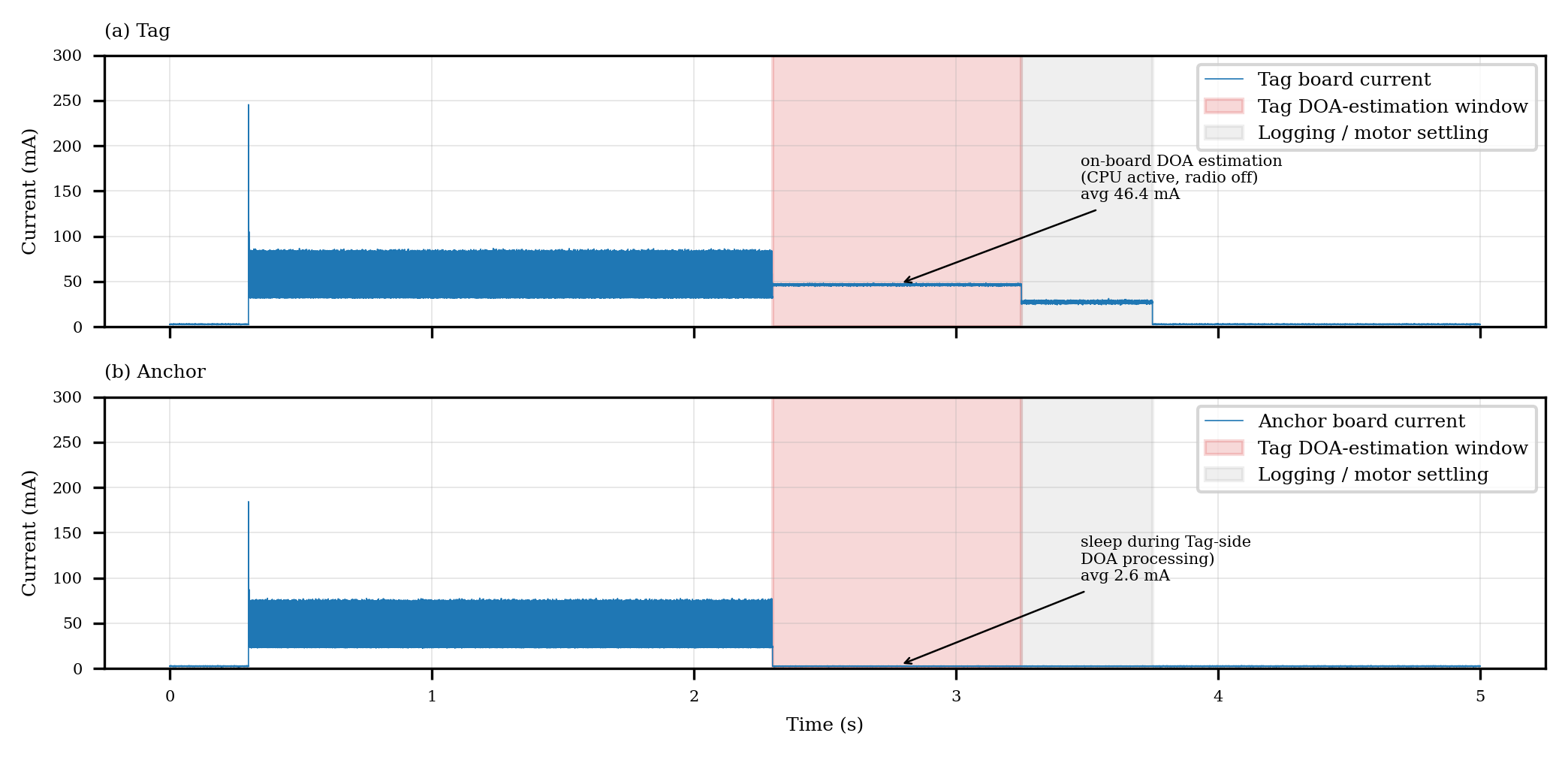}
\caption{Zoomed current trace around the transition from ranging to embedded DOA estimation. The complete measured cycle lasts 39.7 s; only the selected interval containing the DOA computation is displayed for readability.}
\label{fig_tag_anchor}
\end{figure*}
A raw TWR period consists of a request, an anchor reply, and a sleep interval ($T=2$\,ms at 500\,Hz). Averaging $N_{\mathrm{avg}}=128$ exchanges yields one retained range observation. Let $N_{\mathrm{raw}}$ be the number of raw TWR periods in one cycle; the measured ranging-capture phase spans $29.8$\,s at the $500$\,Hz exchange rate in the representative cycle of \autoref{tab:energy_breakdown}. After ranging, the tag executes the embedded \gls{DOA} estimator for a duration $t_{\mathrm{DOA}}$, followed by miscellaneous overhead represented by $t_{\mathrm{extra}}$. The resulting average tag power is
 
\begin{equation}
\resizebox{0.91\columnwidth}{!}{$
P_{\mathrm{tag}} = V\,
\dfrac{N_{\mathrm{raw}}\left(t_{\mathrm{packet}} I_{\mathrm{TX}}
 + t_{\mathrm{reply}} I_{\mathrm{RX}}
 + t_{\mathrm{sleep}} I_{\mathrm{sleep}}\right)
 + t_{\mathrm{DOA}} I_{\mathrm{DOA}}
 + t_{\mathrm{extra}} I_{\mathrm{extra}}}
{N_{\mathrm{raw}} T + t_{\mathrm{DOA}} + t_{\mathrm{extra}}}
$}.
\label{eq:P_tag_corr}
\end{equation}
 
For the anchor, the TX/RX roles are exchanged, and the node is assumed to remain in the selected sleep state during the tag-side \gls{DOA} computation:
 
\begin{equation}
\resizebox{0.96\columnwidth}{!}{$
P_{\mathrm{anchor}} = V\,
\dfrac{N_{\mathrm{raw}}\left(t_{\mathrm{reply}} I_{\mathrm{TX}}
 + t_{\mathrm{packet}} I_{\mathrm{RX}}
 + t_{\mathrm{sleep}} I_{\mathrm{sleep}}\right)
 + t_{\mathrm{DOA}} I_{\mathrm{sleep}}
 + t_{\mathrm{extra}} I_{\mathrm{extra,anchor}}}
{N_{\mathrm{raw}} T + t_{\mathrm{DOA}} + t_{\mathrm{extra}}}
$}.
\label{eq:P_anchor_corr}
\end{equation}
 
Assuming a 300\,mAh battery, the ideal operating lifetime is estimated as
$T_{\mathrm{battery}} = C_{\mathrm{bat}}V/P_{\mathrm{tag}}$, neglecting regulator losses, motor consumption, and battery aging.
 
\begin{figure}[t]
\centering
\includegraphics[width=0.9\linewidth]{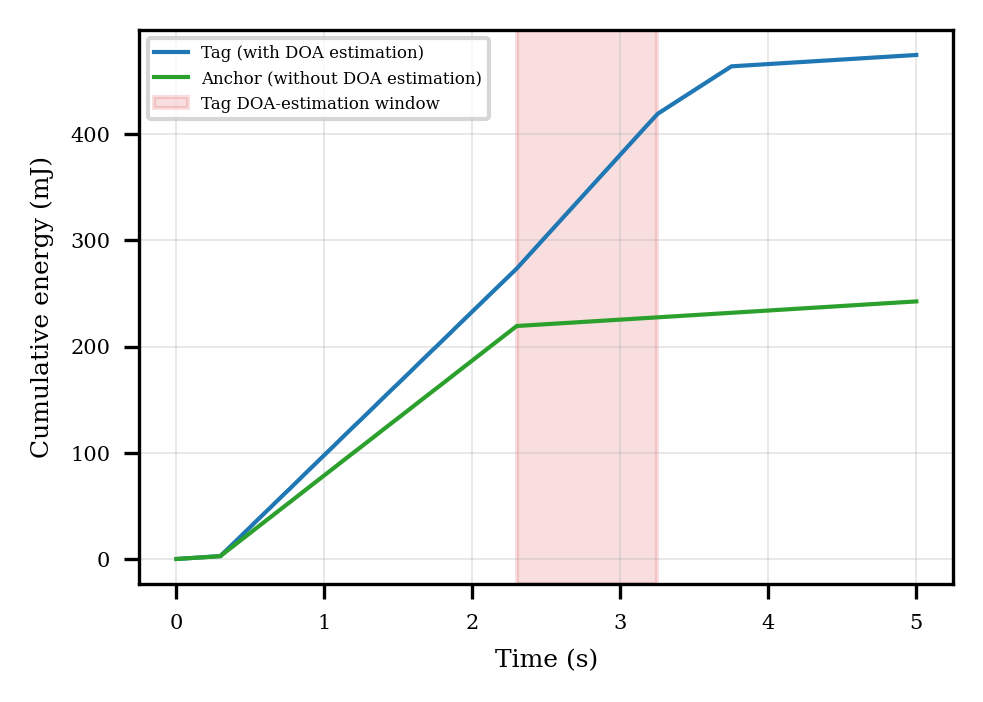}
\caption{Cumulative tag and anchor energy over the zoomed DOA-computation interval. Complete-cycle values (39.7\,s and 4578\,mJ at the tag) are reported in \autoref{tab:energy_breakdown}.}
\label{fig_energy_cmp}
\end{figure}

Power was measured with a Nordic PPK2 at the tag board and isolated SR1020 rail, using GPIO phase markers. \autoref{fig_tag_anchor} and \autoref{fig_energy_cmp} are zoomed around the tag-side \gls{DOA} computation, not the complete 39.7\,s cycle. The tag draws 46.4\,mA during computation, while the anchor sleeps at 2.6\,mA, consistent with \eqref{eq:P_anchor_corr}. As reported in \autoref{tab:energy_breakdown}, computation consumes 144.9\,mJ (about 3\% of the 4578\,mJ cycle); TWR capture dominates the energy budget.

\begin{table}[t]
\centering
\scriptsize
\caption{Measured per-phase board-level energy breakdown of one representative range-domain bearing-estimation cycle on the STM32G473 tag.}
\label{tab:energy_breakdown}
\begin{tabular}{|l|c|c|c|}
\hline
\textbf{Phase} & \textbf{Duration} & \textbf{Energy} & \textbf{Avg.\ current} \\
\hline
TWR ranging capture   & $29.8$\,s  & $4032.6$\,mJ & $41.0$\,mA \\
DOA estimation        & $0.95$\,s  & $144.9$\,mJ  & $46.4$\,mA \\
Logging/motor settling & $3.97$\,s & $356.9$\,mJ  & $27.2$\,mA \\
Inter-trial sleep     & $4.98$\,s  & $43.6$\,mJ   & $2.6$\,mA \\
\hline
\textbf{Total per estimate} & $\mathbf{39.7}$\,\textbf{s} & $\mathbf{4578}$\,\textbf{mJ} & --- \\
\hline
\end{tabular}
\end{table}
 
Finally, the sleep state between \gls{DOA} updates is selected from measured transition energies: a transition from active state $a$ to sleep state $s$ pays off beyond the break-even idle duration
$T_{\mathrm{BE}} = (E_{\mathrm{wake}} + E_{\mathrm{sleep\_entry}})/(P_a - P_s)$, and for a \gls{DOA}-update interval $T_{\mathrm{update}}$ the average system power is
 
\begin{equation}
P_{\mathrm{avg}}(T_{\mathrm{update}}) =
\frac{E_{\mathrm{est}}
+ P_{\mathrm{sleep}}\left(T_{\mathrm{update}}
- T_{\mathrm{cycle}}\right)}{T_{\mathrm{update}}},
\label{eq:p_avg_update}
\end{equation}

where $E_{\mathrm{est}}$ is the measured energy of one complete cycle and $P_{\mathrm{sleep}}$ the measured power of the selected sleep state, quantifying the trade-off between \gls{DOA} refresh rate and battery lifetime.

 \section{Conclusion}
\label{sec:conclusion}
 
This work investigated single-antenna \gls{DOA} estimation using motion-induced virtual apertures for embedded UWB systems. A phase-coherent curvilinear virtual-array MUSIC formulation was developed and evaluated exclusively through numerical simulations, which quantified the effects of SNR, snapshot count, source number, multipath, trajectory curvature, position uncertainty, aperture, and array size.
 
Experimental validation focused on the embedded TWR--IMU range-domain implementation. Using two SR1020 UWB radios and a controlled circular trajectory with a 120~mm antenna lever arm, the estimator was evaluated over 100 consecutive four-revolution sweeps, all of which were retained. Phase-aligned accumulation recovered a 123~mm modulation amplitude, closely matching the measured lever arm. The inferred single-sweep bearing precision was $8.2^\circ$. Because the BNO055 magnetometer exhibited a systematic drift of approximately $2.0^\circ$ per sweep in the motorized metallic setup, these experiments characterize estimator precision rather than absolute world-frame accuracy. The embedded computation required 144.9~mJ per estimate, representing approximately 3\% of the measured cycle energy.
 
These results demonstrate lightweight onboard TWR--IMU bearing estimation without GPS or optical tracking, assuming a known anchor, while identifying magnetometer stability as the main absolute-heading limitation. Future work will address robust inertial/magnetic and RF-phase calibration, embedded phase-coherent snapshot acquisition and MCU MUSIC implementation, and phase-coherent curvilinear-VAA MUSIC under unconstrained motion. A near-field extension will require a spherical-wavefront steering model together with dedicated numerical and experimental validation. Antenna orientation, polarization mismatch, and body blockage will also be investigated.
 
\section*{Acknowledgements}
The authors would like to acknowledge the financial support received from NSERC Alliance (ALLRP~566678-2021), and PROMPT (\#164\_Voix-EERS 2021.06), for the \'ETS-EERS Industrial Research Chair in in-ear technologies, sponsored by EERS Global Technologies~Inc. Furthermore, the authors thank SPARK Microsystems for providing the transceivers used in the experiments.
\IEEEtriggeratref{17}
\bibliographystyle{IEEEtran}
\bibliography{IEEEabrv,References}

@STRING{IEEE_ISCAS     			= "{IEEE} Int. Symp. on Circuits and Syst. ({ISCAS})"}

@STRING{IEEE_VTC              = "{IEEE} Veh. Technol. Conf. ({VTC})"}

@STRING{IEEE_TSP = "{IEEE} Trans. on Signal Process."}

@STRING{IEEE_TCE = "{IEEE} Trans. on Consum. Electron."}

@STRING{IEEE_J_SL = "{IEEE} Sensors Lett."}

@inproceedings{b3,
  author = {B. Liao and S. C. Chan},
  title = {{DOA Estimation of Coherent Signals for Uniform Linear Arrays With Mutual Coupling}},
  booktitle = IEEE_ISCAS,
  pages = {377--380},
  month = {May},
  year = {2011},
  doi = {10.1109/ISCAS.2011.5937581}
}

@article{b6,
  author = {R. Roy and T. Kailath},
  title = {{{ESPRIT} - Estimation of Signal Parameters Via Rotational Invariance Techniques}},
  journal = IEEE_J_ASSP,
  volume = {37},
  number = {7},
  pages = {984--995},
  year = {1989},
  doi = {10.1109/29.32276}
}

@article{b7,
  author = {L. Chen and W. Wei and D. Xia},
  title = {{Adaptive Beamforming Algorithm Based on Residual Neural Networks}},
  journal = {Circuits, Systems, and Signal Process.},
  volume = {44},
  number = {1},
  pages = {556--574},
  month = {January},
  year = {2025},
  doi = {10.1007/s00034-024-02859-z},
  publisher = {Springer}
}

@article{b8,
  author = {M. A. M. Marinho and F. Antreich and S. Caizzone and J. P. C. L. Da Costa and A. Vinel and E. P. De Freitas},
  title = {{Robust Nonlinear Array Interpolation for Direction of Arrival Estimation of Highly Correlated Signals}},
  journal = {Signal Process.},
  volume = {144},
  pages = {19--28},
  month = {March},
  year = {2018},
  doi = {10.1016/j.sigpro.2017.09.025}
}

@article{b9,
  author = {S. Li and L. Li and B. Liu and Y. Song and M. Li and J. Ren and W. Jiang},
  title = {{High-Precision Fast Direction-of-Arrival Estimation Method for Planar Array}},
  journal = {Space: Science \& Technol.},
  volume = {3},
  year = {2023},
  doi = {10.34133/space.0019}
}

@article{b10,
  author = {H. Chen and T. Ballal and T. Y. Al-Naffouri},
  title = {{{DOA} Estimation with Non-Uniform Linear Arrays: A Phase-Difference Projection Approach}},
  journal = IEEE_J_WCOML,
  volume = {10},
  number = {11},
  pages = {2435--2439},
  month = {November},
  year = {2021},
  doi = {10.1109/LWC.2021.3102402}
}

@article{b11,
  author = {W. Si and F. Zeng and C. Hou and Z. Peng},
  title = {{A Sparse-Based Off-Grid {DOA} Estimation Method for Coprime Arrays}},
  journal = {Sensors},
  volume = {18},
  number = {9},
  pages = {3025},
  month = {September},
  year = {2018},
  doi = {10.3390/s18093025}
}

@article{b12,
  author = {Kim, Byung-Ju and Kim, Sangdong and Lee, Jong-Wook},
  title = {A Novel DFT-Based DOA Estimation by a Virtual Array Extension Using Simple Multiplications for FMCW Radar},
  journal = {Sensors},
  volume = {18},
  number = {5},
  pages = {1560},
  year = {2018},
  publisher = {MDPI},
  doi = {10.3390/s18051560}
}

@article{b13,
  author = {F. Quitin and P. De Doncker and F. Horlin and W. P. Tay},
  title = {{Virtual Multiantenna Array for Estimating the Direction of a Transmitter: System, Bounds, and Experimental Results}},
  journal = IEEE_J_VT,
  volume = {67},
  number = {2},
  pages = {1510--1520},
  month = {February},
  year = {2018},
  doi = {10.1109/TVT.2017.2762728}
}

@inproceedings{b14,
  author = {Y. Agram and J. Cheng and F. Quitin},
  title = {{Direction-of-Arrival Estimation Using Virtual Dual-Antenna Receivers: Algorithms and Controlled Experiments}},
  booktitle = IEEE_VTC,
  pages = {1--5},
  month = {June},
  year = {2023},
  doi = {10.1109/VTC2023-Spring57618.2023.10200897}
}

@article{b15,
  author = {{Bou Serhal}, R. E. and Falk, T. H. and Voix, J.},
  title = {{Integration of a Distance Sensitive Wireless Communication Protocol to Hearing Protectors Equipped with In-Ear Microphones}},
  journal = {Meetings Acoust.},
  volume = {19},
  number = {1},
  month = {May},
  year = {2013},
  doi = {10.1121/1.4800452}
}

@inproceedings{b16,
  author = {P. Gupta and S. P. Kar},
  title = {{{MUSIC} and Improved {MUSIC} Algorithm to Estimate Direction of Arrival}},
  booktitle = {Int. Conf. on Commun. and Signal Process. (ICCSP)},
  pages = {757--761},
  month = {April},
  year = {2015},
  doi = {10.1109/ICCSP.2015.7322593}
}

@inproceedings{b17,
  author = {M. W. T. S. Chowdhury and M. Mastora},
  title = {{Performance Analysis of {MUSIC} Algorithm for {DOA} Estimation with Varying {ULA} Parameters}},
  booktitle = {Int. Conf. on Comput. and Inf. Technol. (ICCIT)},
  pages = {1--5},
  month = {December},
  year = {2020},
  doi = {10.1109/ICCIT51783.2020.9392663}
}

@article{b18,
  author = {M. Shaghaghi and S. A. Vorobyov},
  title = {{Subspace Leakage Analysis and Improved {DOA} Estimation With Small Sample Size}},
  journal = IEEE_TSP,
  volume = {63},
  number = {12},
  pages = {3251--3265},
  month = {June},
  year = {2015},
  doi = {10.1109/TSP.2015.2422675}
}

@misc{b22,
  author = {Spark Microsystems},
  title = {{SR10x0 Ranging Application Note (Rev. 0.2)}},
  month = {February},
  year = {2022},
  url = {https://www.sparkmicro.com/wp-content/uploads/2022/03/SR10x0_Ranging_AN_v0.2.pdf}
}

@inproceedings{b23,
  author = {Y. Zhu and S. Xie and M. Yang and M. Zuo},
  title = {{Emitter Localization Using a Single Moving Observer Based on {UKF}}},
  booktitle = {IEEE Int. Conf. Commun. Technol. (ICCT)},
  pages = {1157--1161},
  month = {October},
  year = {2017},
  organization = {IEEE}
}

@misc{b25,
  author = {Spark Microsystems},
  title = {{SR1010 and SR1020 Wireless UWB Transceivers}},
  month = {April},
  year = {2025},
  url = {https://www.sparkmicro.com/product/sr1010-sr1020-transceivers/}
}

@article{b47,
  author = {Zheng, Hang and Zhou, Chengwei and Shi, Zhiguo and Gu, Yujie and Zhang, Yimin D.},
  title = {{Coarray Tensor Direction-of-Arrival Estimation}},
  journal = IEEE_TSP,
  volume = {71},
  issn = {1941-0476},
  pages = {1128--1142},
  year = {2023},
  doi = {10.1109/TSP.2023.3260559}
}

@article{b48,
  title = {{A Hybrid Dynamic Subarray Architecture for Efficient DOA Estimation in THz Ultra-Massive Hybrid MIMO Systems}},
  author = {Tian, Ye and Ren, Jiaji and Wu, Tuo and Liu, Wei and Yuen, Chau and Debbah, M{\'e}rouane and Al-Dhahir, Naofal and Valenti, Matthew C. and So, Hing Cheung and Eldar, Yonina C.},
  journal = IEEE_TSP,
  year = {2026},
  note = {Accepted for publication; preprint available at arXiv:2501.18378},
  url = {https://arxiv.org/abs/2501.18378}
}

@article{b49,
  author = {Abdelbadie, Ali and Mostafa, Mona and Bameri, Salime and Gohary, Ramy H. and Thomas, Dimple},
  title = {{{DoA} Estimation for Hybrid Receivers: Full Spatial Coverage and Successive Refinement}},
  journal = IEEE_TSP,
  volume = {72},
  issn = {1941-0476},
  pages = {4730--4744},
  year = {2024},
  doi = {10.1109/TSP.2024.3459422}
}

@phdthesis{b50,
  author = {Ashok, C.},
  title = {{Efficient Methods for Unambiguous Direction-of-Arrival Estimation with Co-Prime Linear Arrays}},
  school = {Anna University, Chennai, India},
  year = {2022},
  month = dec,
  url = {https://signalprocessingsociety.org/sites/default/files/uploads/newsletter/docs/Ashok_Chandrasekaran_Thesis.pdf}
}

@article{b51,
  author = {Xiao, Disheng and Hu, Shicheng and Kang, Kai and Qian, Hua},
  title = {{An Improved AoA Estimation Algorithm for BLE System in the Presence of Phase Noise}},
  journal = IEEE_TCE,
  volume = {69},
  number = {3},
  pages = {400--407},
  year = {2023},
  publisher = {IEEE},
  doi = {10.1109/TCE.2023.3254595}
}

@ARTICLE{b52,
  author = {Ahmad, Waqas and Hadi, Muhammad Umer and Janjua, Muhammad Bilal and Mumtaz, Shahid},
  journal = IEEE_TCE,
  title = {{A Novel Self-Calibrated UWB-Based Indoor Localization System for Consumer Electronics Applications}},
  year = {2024},
  volume = {70},
  number = {1},
  pages = {1672--1684},
  doi = {10.1109/TCE.2024.3369193}
}

@ARTICLE{b53,
  author = {Khan, Adeel and Ullah, Imran and Akbar, Muhammad},
  journal = IEEE_TCE,
  title = {{An Optimized Location-Based System for the Improvement of Consumer Services}},
  year = {2024},
  volume = {70},
  number = {2},
  pages = {4998--5006},
  doi = {10.1109/TCE.2024.3351280}
}

@article{Sabbar2021,
  author  = {Salman, Fatimah A. and Sabbar, Bayan M.},
  title   = {Coprime Array Parameters Optimization for {DOA} Estimation},
  journal = {Iraqi Journal of Information and Communication Technology},
  volume  = {4},
  number  = {2},
  pages   = {23--32},
  year    = {2021},
  doi     = {10.31987/ijict.4.2.131}
}

@article{capon1969,
  author  = {Capon, Jack},
  title   = {High-Resolution Frequency-Wavenumber Spectrum Analysis},
  journal = {Proceedings of the IEEE},
  volume  = {57},
  number  = {8},
  pages   = {1408--1418},
  month   = aug,
  year    = {1969},
  doi     = {10.1109/PROC.1969.7278}
}

@book{vantrees2002,
  author    = {Van Trees, Harry L.},
  title     = {Optimum Array Processing: Part IV of Detection, Estimation,
               and Modulation Theory},
  publisher = {Wiley},
  address   = {New York, NY, USA},
  year      = {2002},
  volume    = {1},
  series    = {Detection, Estimation, and Modulation Theory},
  isbn      = {978-0-471-09390-9}
}

@article{liu2018tap,
  author  = {Liu, Zhang-Meng and Zhang, Chenwei and Yu, Philip S.},
  title   = {{Direction-of-Arrival Estimation Based on Deep Neural Networks With Robustness to Array Imperfections}},
  journal = {IEEE Transactions on Antennas and Propagation},
  volume  = {66},
  number  = {12},
  pages   = {7315--7327},
  month   = dec,
  year    = {2018},
  doi     = {10.1109/TAP.2018.2874430}
}

@article{elbir2020deepmusic,
  author  = {Elbir, Ahmet M.},
  title   = {{DeepMUSIC: Multiple Signal Classification via Deep Learning}},
  journal = IEEE_J_SL,
  volume  = {4},
  number  = {4},
  pages   = {1--4},
  month   = apr,
  year    = {2020},
  doi     = {10.1109/LSENS.2020.2980384}
}

@article{zhao2022cnn,
  author  = {Zhao, Fangzheng and Hu, Guoping and Zhan, Chenghong and Zhang, Yule},
  title   = {{DOA Estimation Method Based on Improved Deep Convolutional Neural Network}},
  journal = {Sensors},
  volume  = {22},
  number  = {4},
  pages   = {1305},
  year    = {2022},
  doi     = {10.3390/s22041305},
  publisher = {MDPI}
}
 
\end{document}